\documentclass[useAMS,usenatbib]{mn2e}
\usepackage{graphicx}
\usepackage{graphics}
\usepackage{natbib}
\usepackage{amssymb}
\usepackage{aas_macros}

\begin{document}

\title[Compact groups in theory and practice - II ]{Compact groups in
theory and practice - II. Comparing the observed and predicted nature
of galaxies in compact groups.}

\author[Brasseur et al.] {Crystal M. Brasseur$^1$, Alan W. McConnachie$^{1,2}$, Sara L. Ellison$^1$,\newauthor David R. Patton$^{3,4}$\\$^1$Department of Physics \& Astronomy, University of Victoria, Victoria, B.C. V8P 1A1, Canada\\$^2$Herzberg Institute of Astrophysics, National Research Council, Victoria, BC V9E 2E7, Canada\\$^3$Department of Physics \& Astronomy, Trent University, 1600 West Bank Drive, Peterborough, ON K9J 7B8, Canada\\$^4$Visiting Researcher, Department of Physics \& Astronomy, University of Victoria, Victoria, B.C. V8P 1A1, Canada}

\maketitle

\begin{abstract}
We examine the properties of galaxies in compact groups identified in
a mock galaxy catalogue based upon the Millennium Run simulation.  
The overall properties of groups identified in projection are in general agreement with
the best available observational constraints.
However, only \mbox{$\sim 30$\,\%} of these
simulated groups are found to be truly compact in 3 dimensions,
suggesting that interlopers strongly affect our observed understanding
of the properties of galaxies in compact groups.  These simulations
predict that genuine compact group galaxies are an extremely
homogeneous population, confined nearly exclusively to the red
sequence: they are best described as ``red and dead'' ellipticals.
When interlopers are included, the population becomes much more
heterogeneous, due to bluer, star-forming, gas-rich, late-type
galaxies incorrectly identified as compact group members. These models
suggest that selection of members by redshift, such that the
line-of-sight velocity dispersion of the group is less than
1000\,km\,s$^{-1}$, significantly reduces contamination to the 30\%
level. Selection of members by galaxy colour, a technique used
frequently for galaxy clusters, is also predicted to dramatically
reduce contamination rates for compact group studies.  

\end{abstract}


\section{Introduction}
\label{Introduction}

Compact groups (CGs) of galaxies are spatially dense groupings of
generally 4 or more galaxies. In contrast to galaxies found in cluster
cores (\citealt{Z80}), the projected velocity dispersions of CGs are
generally small (\mbox{$\sim 200$\,km\,s$^{-1}$;} \citealt{HMH92}).
This combination of low velocity dispersions and high spatial
densities makes CGs an optimal environment for dynamical interactions
and mergers to occur.  Galaxy interactions, whether in groups or
pairs, have been shown to influence the observed properties of
galaxies (\citealt{LT1978}; \citealt{K1987}; \citealt{SD2000};
\citealt{Barton2000} \citealt{Lam03}; \citealt{Alonso2004};
\citealt{Nik04}; \citealt{Pat05}; \citealt{Gell06}), and so CGs are an
ideal environment in which to study the effects of encounters on
galaxy morphology.

Observational studies of CGs reveal that galaxies in these
environments tend to be systematically different from the field
population. Perhaps most striking of all, the fraction of early-type
galaxies in CGs in magnitude-limited surveys is significantly higher
than in the field; \citet{H88} found 51$\%$ of their CG galaxies were
early-type, compared to \mbox{$\sim 20$\,\%} of the field
(\citealt{N73,G80}). Similarly, \citet{P95} found 55$\%$ of galaxies
in their CG sample were early-type. Early-type galaxies can be formed
via the merging of late-type systems (eg. \citealt{B89}), and so these
results can be explained by frequent interactions and mergers between
the galaxies in the CG environment.

One result of mergers which might be expected is an enhanced star
formation rate (SFR), triggered by the re-distribution of gas as
galaxies tidally interact.  Close pairs of galaxies are observed to
have enhanced SFRs, consistent with this picture (e.g., Carlberg
et. al. 1994, \citealt{Ellison2008} and references therein) . Yet in
clusters, SFRs are low when the local density is high
(e.g. \citealt{Gomez2003}, \citealt{Lewis2002}), therefore it is not
clear \emph{a priori} which way the SFRs will go in compact group
galaxies.  %Observations by \citet{M94} and \citet{IP1999} have shown
%that CGs galaxies have SFRs similar to those of isolated galaxies.
One possible scenario is that galaxy interactions led to an
exhaustion of gas in these galaxies through previous starbursts and/or
stripping (either tidal or ram pressure). Indeed, \citet{WR87} found
CG galaxies to be deficient in HI by a factor of two on average when
compared to galaxies in loose groups, and \citet{M95} found a
significant deficiency in total radio emission from CG spirals
compared to field spirals, again implying a reduced gaseous content.

Differences have also been measured in the mean colours of CG and
field galaxies.  In very broad terms, redder galaxies in the field
generally consist of an older and/or more metal rich stellar
population than bluer galaxies, although other effects such as
extinction can play a significant role. \citet{L04} found the
rest-frame colours of their CG sample were on average redder than
field galaxies at approximately the $2\sigma$ level.  \citet{D07} also
found that the mean colours of galaxies in CGs were redder and had a
smaller dispersion in colour than galaxies in their control sample.

Clearly, over the past several decades there has been considerable
observational effort to determine the properties of galaxies in
CGs. Cosmological simulations now have sufficient detail and
statistics to examine the properties of galaxies in CGs and provide a
meaningful comparison with observations. For example, do these models
predict that compact groups should have an excess of early-type
galaxies and, if so, how large is the excess? How does this compare to
observations? What about the colours and star formation rates of the
galaxies? Can any differences between the observed and predicted
properties of CG galaxies be traced back to inadequacies in the
modeling or observational effects?

In \citet{M07} (hereafter Paper~I), we identified compact groups in a
mock galaxy catalogue (\citealt{B05}) based upon the Millennium Run
simulation (\citealt{S2005}) and investigated their spatial
properties. In this paper, the second in the series, we determine the
main physical properties of the galaxies in these CGs and compare them
to observational results to determine if our observational and
theoretical understanding of these galaxies are in conflict and, if
so, why.

In Section~2, we review how CGs are identified and quantified from the
mock catalogue. We also review how the main physical parameters for
the simulated galaxies, which we shall compare to observations, are
calculated, and discuss possible limitations. In Section~3, we
determine the range of properties shown by the CGs in the mock
catalogue and compare them to a control sample. We postpone the
majority of the discussion of our results and comparison to
observations until Section~4, and we summarise our results in
Section~5.

\section{Sample Selection}
\label{SS}

In Paper~I, CGs are identified in a mock galaxy catalogue based upon
their projected characteristics using the original Hickson criteria
(\citealt{H82}). We refer to these galaxy associations identified in
the simulation as Hickson Associations (HAs), in recognition of the
fact that we find \mbox{$\sim 70$\,\%} of HAs in the simulation are
not truly compact in three dimensions (for example, they are
projections of looser groups or physically unassociated galaxies) and
that they may contain interlopers. Compact Associations (CAs) are
defined as being that subset of HAs which are truly compact in three
dimensions. In Paper~I, the degree of compactness of each group is
quantified using the concept of three dimensional {\it linking
length}, $\ell$ (eg. \citealt{HG1982}). We show that a good criterion
for defining CAs in simulations, where 3D information is available,
are those HAs which have $\ell < $ 200~h$^{-1}$ kpc. `CGs' refers to
observationally identified systems, and `HCGs' refers explicitly to
those systems identified in the original \citet{H82} catalogue.
Table~1 summarizes this terminology.

\begin{table}
\begin{center}
\caption{Summary of terminology used in this paper.}
\begin{tabular}{ccl}
\hline
System &     Acronym &   Definition \\
\hline
(Hickson) Compact Group&    (H)CG    &   Observationally \\&&identified systems \\&&(using the original 
\\&&Hickson (1982) \\
\vspace{0.2cm} &&criteria).  \\

Hickson Associations&       HA     &  Identified in the \\
&&mock catalogue using \\
\vspace{0.2cm} 
&&Hickson's criteria.   \\ 

Compact Associations&    CA       &  The subset of HAs\\&&which are truly 
\\&&compact in \\
&&three dimensions.  \\ 
\hline
\end{tabular}
\end{center}
\end{table}

\subsection{The identification of compact groups}

The Millennium Run simulation by \cite{S2005} evolves 2160$^3$ dark
matter particles in a cube with 500 h$^{-1}$ Mpc sides, assuming a
$\Lambda$ Cold Dark Matter (CDM) cosmology ($h$ = 0.73,
$\Omega_{\Lambda}$=0.75, $\Omega_{M}$=0.25), where the Hubble constant
is parameterized as $H_0$=100\,$h$\,km\,s$^{-1}$Mpc$^{-1}$. \citet{LB07}
track the formation of galaxies from the dark matter distribution
using semi-analytic techniques, and their resulting catalogue provides
the 3D positions and velocities of the galaxies, and various
properties such as bulge and stellar masses, colours and SFRs.

Mock galaxy catalogues were created from the output of the
\citet{LB07} catalogue using the Mock Map Facility (MoMaF) code of
\citet{B05}.  We use the ``Blaizot\_Allsky\_PT\_1'' catalogue which is
publicly available on the Millennium
website\footnote{http://www.mpa-garching.mpg.de/millennium/}.  This
mock catalogue contains $\sim$ 5.7 million galaxies brighter than $m_r
= 18$. Paper~I applies the Hickson criteria to this mock galaxy
catalogue and examines the spatial properties of the HAs identified.

In Paper~I, 15\,122 HAs were identified in the mock galaxy catalogue,
consisting of a total of 64\,525 galaxies. However, only 28\% of these
HAs are CAs (ie. physically compact with no interlopers). This
strongly suggests that interlopers have had a significant effect on
the implied observational properties of galaxies in CGs, where it is
more difficult to determine the three dimensional reality of the
systems being studied.

We emphasise that the term `interloper' does not necessarily imply
that the galaxy is at a significantly different redshift to the CG;
rather, it means that the galaxy is unlikely to have evolved in the
dense environment expected of a CG. We define CAs as that subset of HAs with $\ell < $ 200~h$^{-1}$
kpc. Any galaxy greater than 200~h$^{-1}$ kpc away from the main
concentration of galaxies of a CG is therefore classed as an
interloper, even though it may belong to the same larger-scale
environment as the CG (eg. a surrounding loose group) and may not have
a discordant redshift. Nevertheless, its inclusion in a study of the
properties of CG galaxies - which we are ultimately interested in
because of their evolution in an environment where interactions should
be common - could potentially bias any conclusions which are drawn. In
addition, the grouping of galaxies which remains should not strictly
be classed as a CG, since it probably would not satisfy the selection
criteria for a CG (particularly regarding the number of members and
the surface brightness criteria) were it not for the presence of the
interloper(s).

\subsection{Summary of relevant semi-analytics within the simulation}

In the semi-analytic models on which the mock galaxy catalogues are
based, physically motivated recipes for the baryonic distribution are
applied to the collisionless dark matter distribution to create
realistic galaxy populations. In very broad terms, galaxy formation in
these models is the process by which hot gas cools within dark matter
haloes to form stars. A galaxy's merger history, its star formation
rate and various feedback processes combine to determine its
evolution. Galaxies are centered at the position of the most bound
particle in the dark matter halo. Due to the high resolution of the
Millennium Run, the positions and velocities of these galaxies can be
accurately determined by following the orbits and merging histories of
the (sub-)haloes within the simulation.

Below is a brief overview of the physical treatment of various
processes which govern the main observable properties of the galaxies
which we will examine in detail in Section~3. We refer the reader to
\citet{LB07}, and references therein, for a more complete description
of the semi-analytic methods. The initial identification of the
compact groups from these semi-analytic catalogues was made in Paper~I
using the observational criteria of \cite{H82}. Recently, these mock
catalogues have been used in a similar way for a similar purpose by
\cite{M08} in their study of close galaxy pairs.

\subsubsection{Cold gas}

\cite{LB07} assume that as each dark matter halo collapses, some
baryonic matter collapses with it (determined as a fraction of the
dark matter mass). This gas is shock heated to the virial temperature
of the halo and, following \citet{K99} and \citet{S2001}, a cooling
time is computed for the gas using cooling curves dependent on the
temperature and metallicity (\citealt{SD93}). A cooling radius,
defined as the radius at which the cooling time of a galaxy is equal
to the age of the Universe, is calculated, and all gas within the
cooling radius is then treated as cold gas\footnote{The cooling radius
will continue to propagate outwards with time in a galaxy until either
all of the hot gas in the halo has cooled or more material is injected
into the halo via feedback processes or mergers with other
galaxies.}. The cold gas is then accreted onto the disc of the central galaxy
in the halo on the free-fall timescale.  There is assumed to be no cooling onto
 satellite galaxies.

While the detailed chemistry of the cold gas cannot be followed in the
simulation, we assume that the cold gas fraction will trace the HI
abundance. This allows comparison to observations and seems a
reasonable assumption, considering the fraction of other atomic or
molecular gases compared to HI in galaxies is usually small.

\subsubsection{Star formation}

Star formation occurs in two modes. In the first, as long as the mass
of cold gas is greater than a critical value, $M_{crit}$, star
formation proceeds at a continual rate defined as

\begin{equation}
\dot{M_*}=\frac{\alpha (M_{cold}-M_{crit})}{t_{dyn}}~.
\label{sfr_eq}
\end{equation}

\noindent $M_{cold}$ and $t_{dyn}$ are the cold gas mass and the
galactic dynamical time, respectively, and $\alpha$ is the efficiency
by which gas is converted into stars. In the second mode, star
formation occurs in starbursts triggered by mergers
(Section~\ref{Bulge_mass_fraction}).

\subsubsection{Luminosity and colour}

Photometric properties of the galaxies are determined using the
stellar population synthesis models of \citet{BC2003}.  An initial
mass function is adopted and used to compute the number of stars
formed in each mass interval. These stars then evolve along
theoretical evolutionary tracks, primarily governed by their mass and
metallicity.  The spectral energy distribution (SED) of the galaxy is
computed by convolving the evolution of the SED of each
single-age stellar population with the star formation history of the
galaxy.  Convolving the SED with the filter responses gives the
luminosities and colours of each galaxy.

\subsubsection{Bulge mass fraction}
\label{Bulge_mass_fraction}

Galaxy bulges form via mergers and disc instabilities. Discs are
stable if

\begin{equation}
\frac{V_c}{\sqrt{GM_{disc}/r_{disc}}} \geq 1~.
\label{bulge_eq}
\end{equation}

\noindent $M_{disc}$, $r_{disc}$ and $V_c$ are the mass, radius and
rotational velocity of the disc, respectively.  The ratio on the left
hand side of Equation~(\ref{bulge_eq}) is computed for every galaxy at
every time-step. If necessary, stellar mass is transferred from the
disc to the bulge until the inequality is satisfied.

\begin{figure*}
  \begin{center}
    \includegraphics[width=84mm]{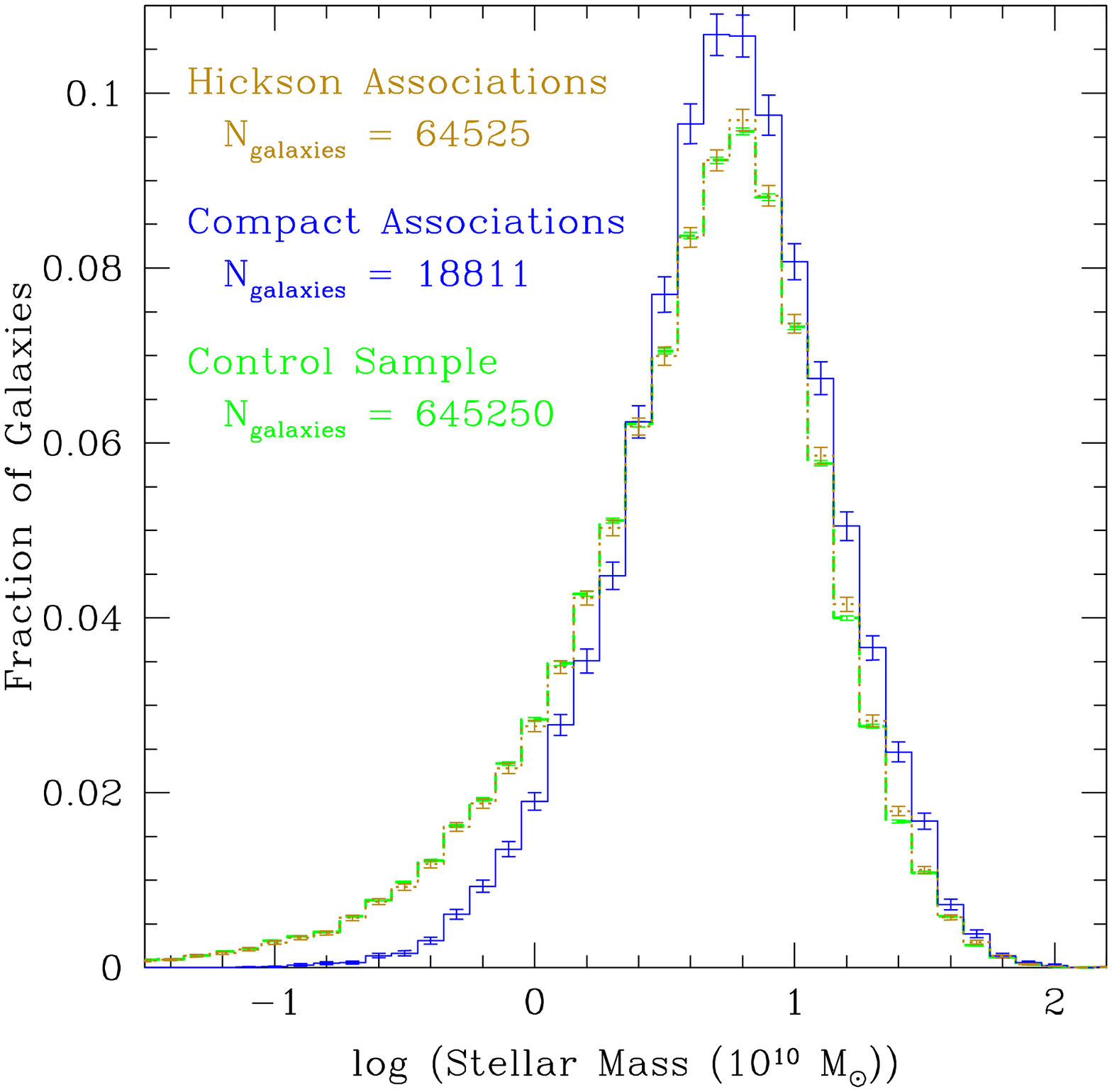}
    \includegraphics[width=84mm]{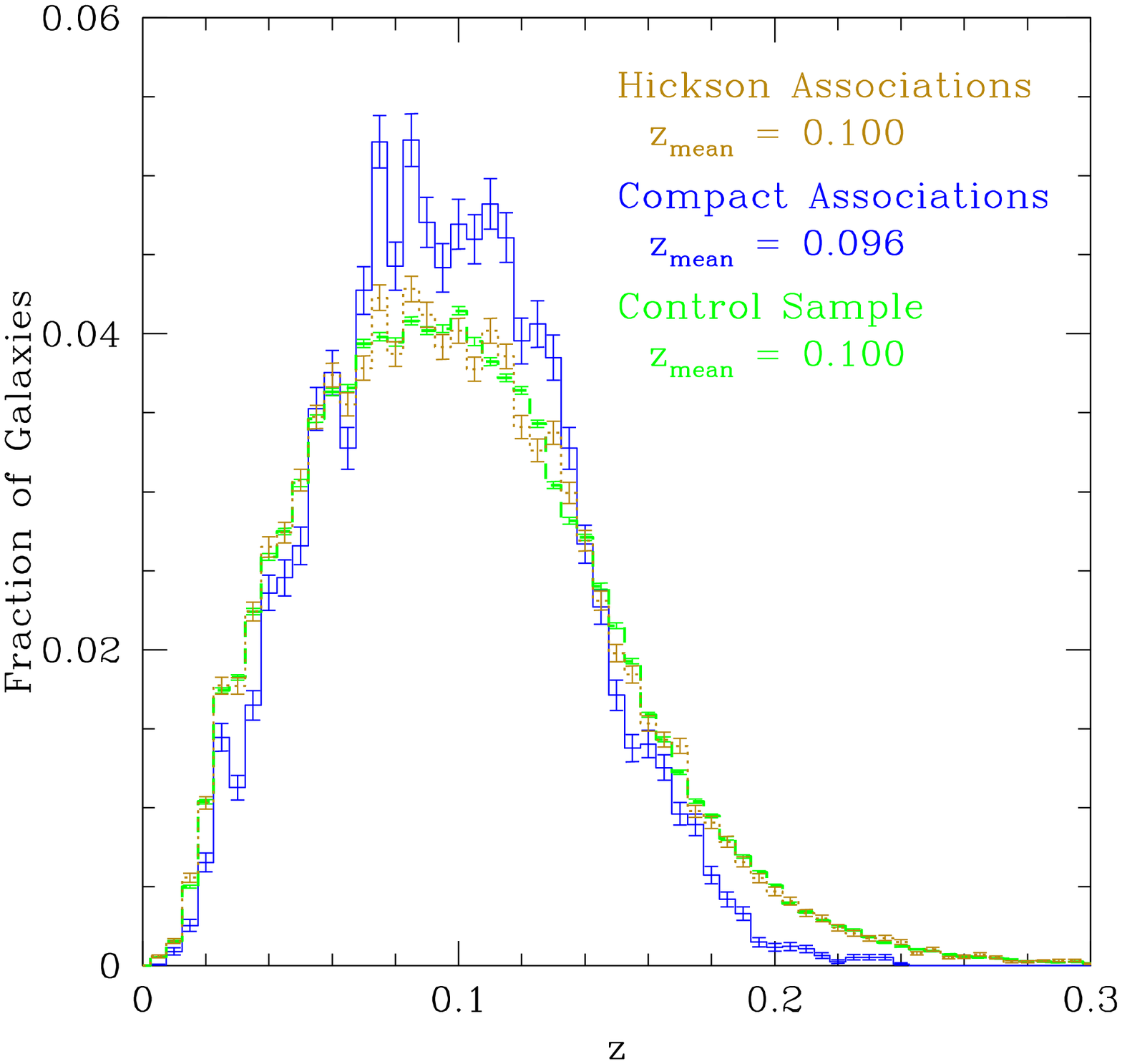}
    \caption{Left panel: Stellar mass distribution for galaxies in
    Hickson Associations (HAs) is shown in dotted brown, the subset of
    these which are compact in three dimensions and contain no
    interlopers (ie. Compact Associations, CAs) is show in solid blue,
    and the control sample is shown in dashed green. The histograms
    have been normalized to have the same total number of galaxies in
    each sample.  Right panel: Redshift distribution of our sample
    galaxies using same line styles as the left panel.  The control
    sample was constructed such as to be matched in both stellar mass
    and redshift to the HA galaxies. One dimensional
    Kolmogrov$-$Smirnov (KS) tests between the HA galaxies and the
    control sample galaxies show that for both the stellar mass and
    redshift distributions, the null hypothesis is accepted at the
    $>90\,\%$ level. }
    \label{stellarmassHCG}
  \end{center}
\end{figure*}

For mergers between galaxies, the `collisional starburst' method of
\citet{SPF01} is applied.  If two galaxies, $G_1$ and $G_2$, with
baryonic masses $M_1 > M_2$, merge, the gas contained within both
galaxies coalesce to form the disc of the post-merger galaxy,
$G_{f}$. The bulge of $G_f$ is composed of the bulge stars from $G_1$
and all the stars from $G_2$.

During a merger where $M_1 >> M_2$, no star formation is triggered.
In the case of a major merger ($M_2/M_1 > 0.3$), the discs are
completely destroyed and all the stellar mass is transferred to the
bulge component of $G_{f}$\footnote{It is possible for $G_f$ to form a
disc at a later time by the accretion of cold gas.}.  Additionally, a
fraction of cold gas from the two merging galaxies is instantaneously
consumed in a starburst. These new stars are added to the bulge
component of $G_{f}$.
 
\section{The properties of galaxies in compact groups}
\label{the_impact_of_projection_effects}

In this section, we determine how the properties of galaxies in HAs
(the equivalent of HCGs in the mock catalogue) compare to the field
population. Further, we compare the properties of CAs to those of
HAs, to gain insight into how observational studies of CGs may have
been biased by the presence of a significant number of interlopers. We
note that, since CAs are an interloper-free subset of HAs, they do not
have an exact observational analogue with which to compare directly.

\subsection{The control sample}

Stellar mass and redshift both correlate with many physical properties
of galaxies due to dynamical and evolutionary considerations.  We
therefore compare the properties of the galaxies in our sample to
those in a field (control) sample selected to match the mass and
redshift distribution of HAs. Differences
between the field and HAs can then be connected to their different
environments, not to a difference in the stellar mass or redshift.  We
match the control sample to HAs because we will first compare HAs to
the control in order to make a comparison analogous to observational
comparisons of CGs and field galaxies. Later on, we compare HA to CA
galaxies to understand the impact of contamination, therefore no
control is required to match CA galaxies.

A control sample was constructed from the initial field sample which
consisted of all galaxies in the Blaizot catalogue which are not
members of a HA. For each HA galaxy, we found 10 unique field galaxies
which have the same stellar mass and redshift as the HA galaxy to
within a tolerance of 10\,\%. The control sample contains 645\,250
galaxies and its stellar mass and redshift distributions are shown in
Figure~\ref{stellarmassHCG}. The left panel of
Figure~\ref{stellarmassHCG} shows the stellar mass distribution of all
galaxies in HAs (dotted brown), CAs (solid blue) and our control
sample (dashed green), where each has been normalised to have the same
total number of galaxies. The distributions for HAs and the control
sample effectively lie on top of one another. The right panel shows
the equivalent distributions for the redshift distribution of the
galaxies in the groups. One dimensional Kolmogrov-Smirnov (KS) tests
between the HA galaxies and the control sample galaxies show that for
both mass and redshift, the null hypothesis that both distributions
are drawn from the same underlying distribution is acceptable at the
$>90\,\%$ level. Galaxies in CAs have a slightly higher mean stellar
mass than galaxies in HAs (${\bar M_*}\simeq 7.9 \times 10^{10}
M_{\odot}$ compared to ${\bar M_*} \simeq 6.8 \times 10^{10}
M_{\odot}$) and a slightly lower mean redshift (${\bar z}_{CA} =
0.096$ compared to ${\bar z}_{control} = 0.100$).

\subsection{Colour}
\label{Colour}

\begin{figure}
  \begin{center}
    \includegraphics[width=84mm]{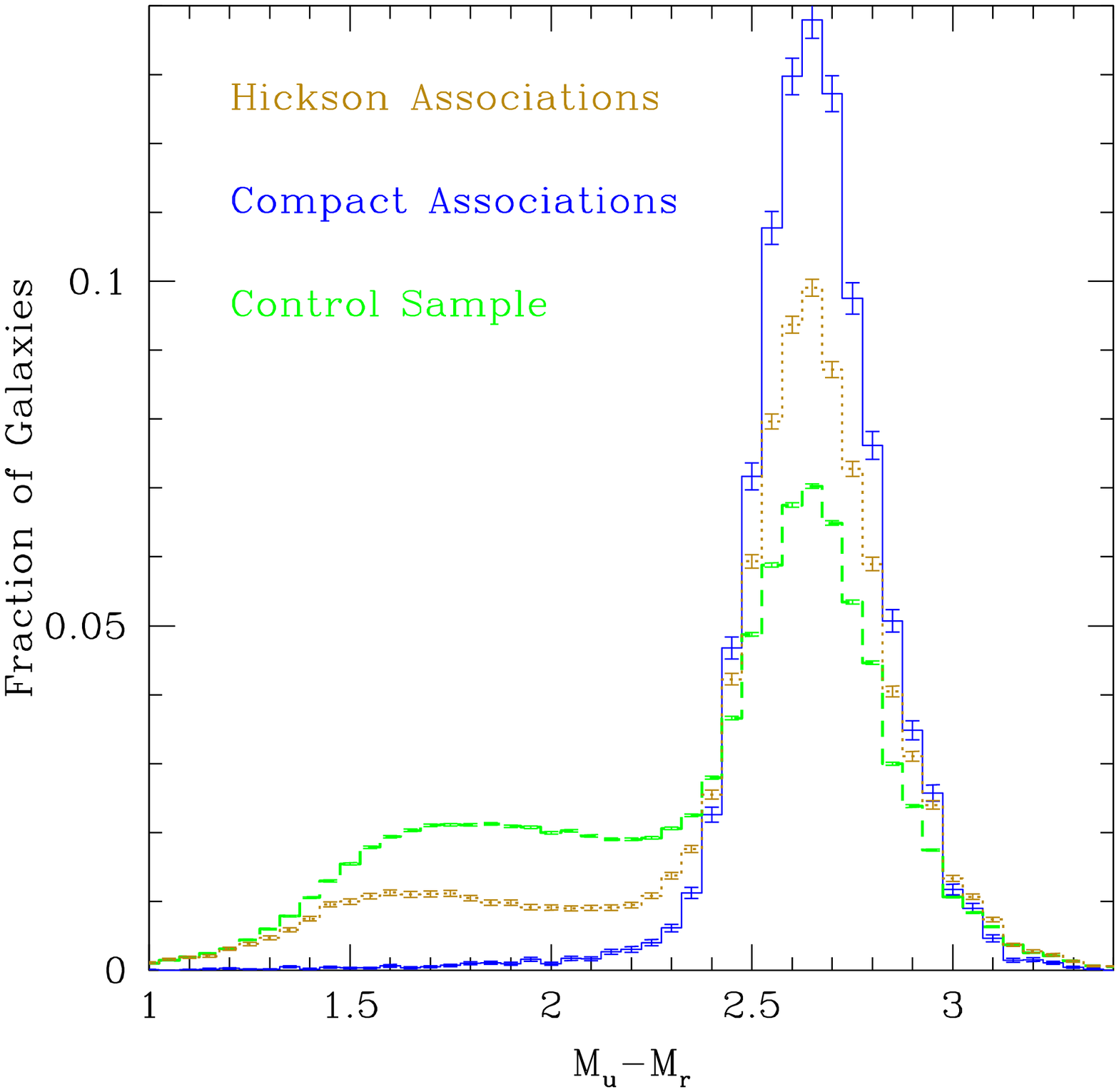}
    \caption{$\left(M_{u}-M_{r}\right)$ colour distributions for
      galaxies in HAs, CAs and our control sample, with the same line
      styles and scaling as used in Figure~\ref{stellarmassHCG}. The distribution of the
      control sample is bi-modal, representing galaxies in the red
      sequence and blue cloud. HAs are also bi-modal, but there is a
      far greater proportion of redder galaxies in comparison to the
      control sample. In contrast, CAs are nearly all red,
      $\left(M_{u}-M_{r}\right) \gtrsim 2.25$, with only a very low
      level tail to bluer colours.}
    \label{u_r_HCG}
  \end{center}
\end{figure}

In observational studies, galaxies in CGs are found on average to be
redder than the field population (\citealt{L04}; \citealt{D07}).
Galaxy colour is a broad indication of the age and metallicity of the
luminosity-weighted mean stellar populations of a galaxy, with older
populations and more metal-rich populations generally appearing
redder. To some extent, colour also correlates with galactic
morphology, with late-type, spiral galaxies generally appearing bluer
than early-type, elliptical galaxies.

For colour analysis in this paper, we use the rest frame Sloan filters
({\it ugriz}) available in the mock catalogue which take into account
the effects of dust in each galaxy. Figure~\ref{u_r_HCG} shows the
$\left(M_{u}-M_{r}\right)$ colour distribution of galaxies in the HAs,
CAs and the control sample, represented by the equivalent line styles
and scaling to Figure~\ref{stellarmassHCG}.

The control sample distribution is bi-modal, with the largest and
reddest peak centered near $\left(M_{u}-M_{r}\right) \simeq 2.7$ and a
broader, secondary, bluer feature centered around
$\left(M_{u}-M_{r}\right) \simeq 1.8$. The $\left(M_{u}-M_{r}\right)$
colour distribution of galaxies in HAs is qualitatively similar to the
control sample, showing a similar bi-modal structure. However, the red
peak is much more dominant for the HAs than for the control, and the
bluer feature is considerably weaker. In all, 79\,\% of galaxies in
HAs are redder than $\left(M_u - M_r\right) = 2.25$, compared to
63\,\% for the control.

The distribution of CA galaxies in Figure~\ref{u_r_HCG} stands out in
comparison to the control and HA galaxies; in contrast to these two
populations, virtually no galaxies in CAs are bluer than
$\left(M_{u}-M_{r}\right) \sim 2.25$ and the distribution is clearly
uni-modal. Given that the contamination by the control sample consists
of both blue and red galaxies, HAs appear to have a population of blue
galaxies which are mistaken for true CG galaxies. However, in the
environment of CAs (ie. all galaxies are in very close proximity),
there appears to be a dearth of blue and a dominance of red galaxies.
Recent work by \cite{Baldry2006} has also found that semi-analytical models 
predict the fraction of red galaxies as a function of environment and mass that is qualitatively similar to observations.

\subsection{Early- and late-type galaxies}

\citet{D80} was the first to show that the proportion of elliptical
and lenticular galaxies increases at the expense of spiral galaxies in
regions with higher projected galaxy density.  Numerical simulations
(\citealt{RN79}; \citealt{FS82}; \citealt{B92}) support this result by
demonstrating that a merger between two spiral galaxies almost always
ends in an elliptical galaxy. Therefore, environments rich in
interactions and merger events (that is, high galaxy density and low
velocity dispersions) are expected to be fertile regions for the
formation of elliptical galaxies.

The above expectations appear to have been observationally confirmed
for CGs.  Studies have shown that CGs contain a significantly
different morphological population of galaxies than does the field
(\citealt{H82}; \citealt{WR87}; \citealt{Sulentic1987}; \citealt{H88};
\citealt{R89}; \citealt{P94}; \citealt{L04}; \citealt{AC07}).
\citet{H88} have shown that compact groups contain a higher fraction
of early-type galaxies as compared to a field population and more
recently, \citet{AC07} show that compact groups isolated from
large-scale structures have a higher fraction of S0 galaxies as
compared to the field.

We quantify the fraction of early- and late-type galaxies in our mock
survey by using the ratio of bulge-to-total stellar mass ($B/T$) for
each galaxy. Following \cite{Sc1996}, we consider a galaxy to be
bulge-dominated (early type) if its bulge-to-total ratio is $ B/T
\gtrsim 0.6$ and disc-dominated (late-type) if $ B/T \lesssim
0.4$. However, these cuts are indicative only, and we stress that the
$B/T$ ratio is not sufficient by itself to determine morphology from
an observational perspective, although it is the most analogous
quantity in the simulations and is still useful from a statistical
perspective.

\begin{figure*}
  \begin{center}
    \includegraphics[width=84mm]{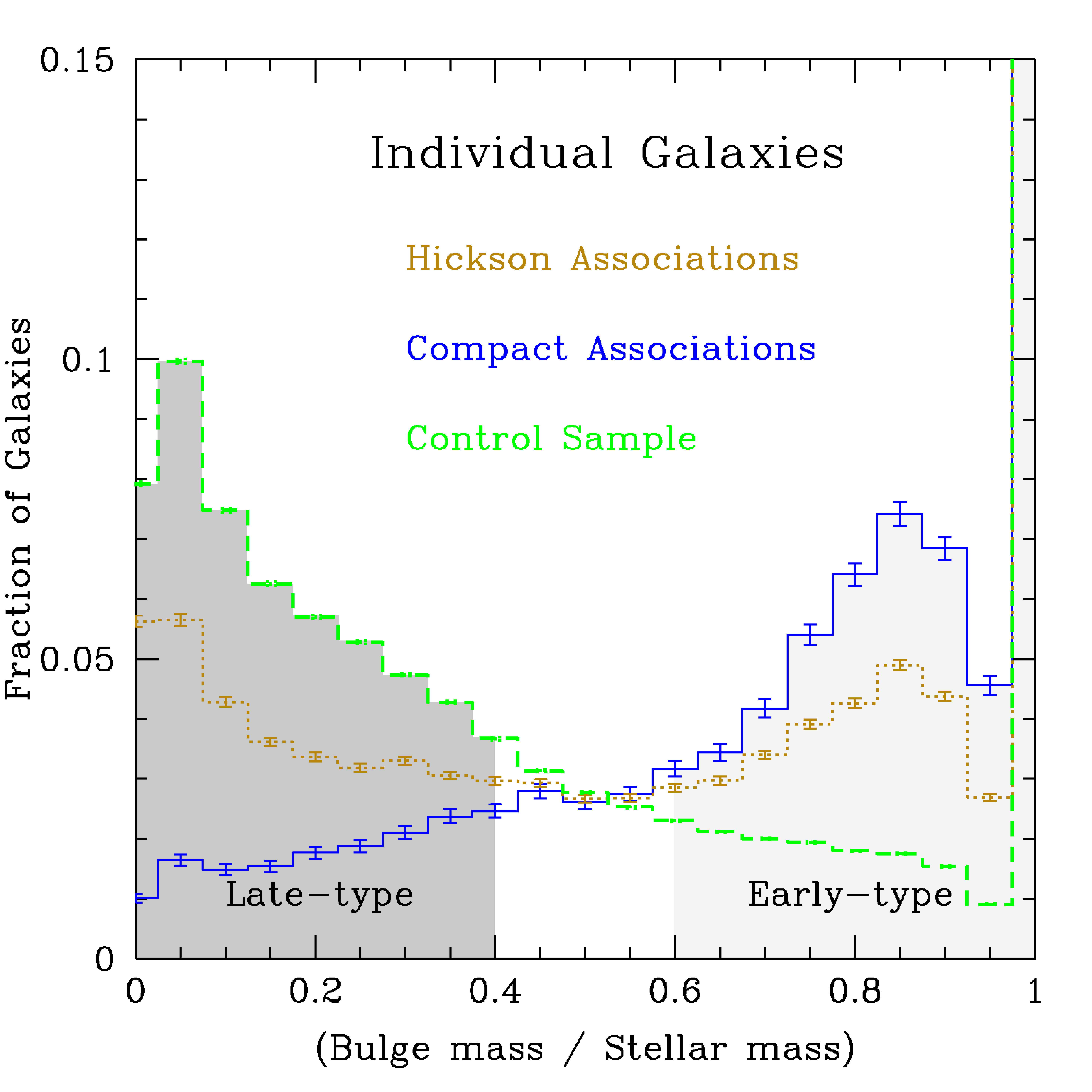}
    \includegraphics[width=84mm]{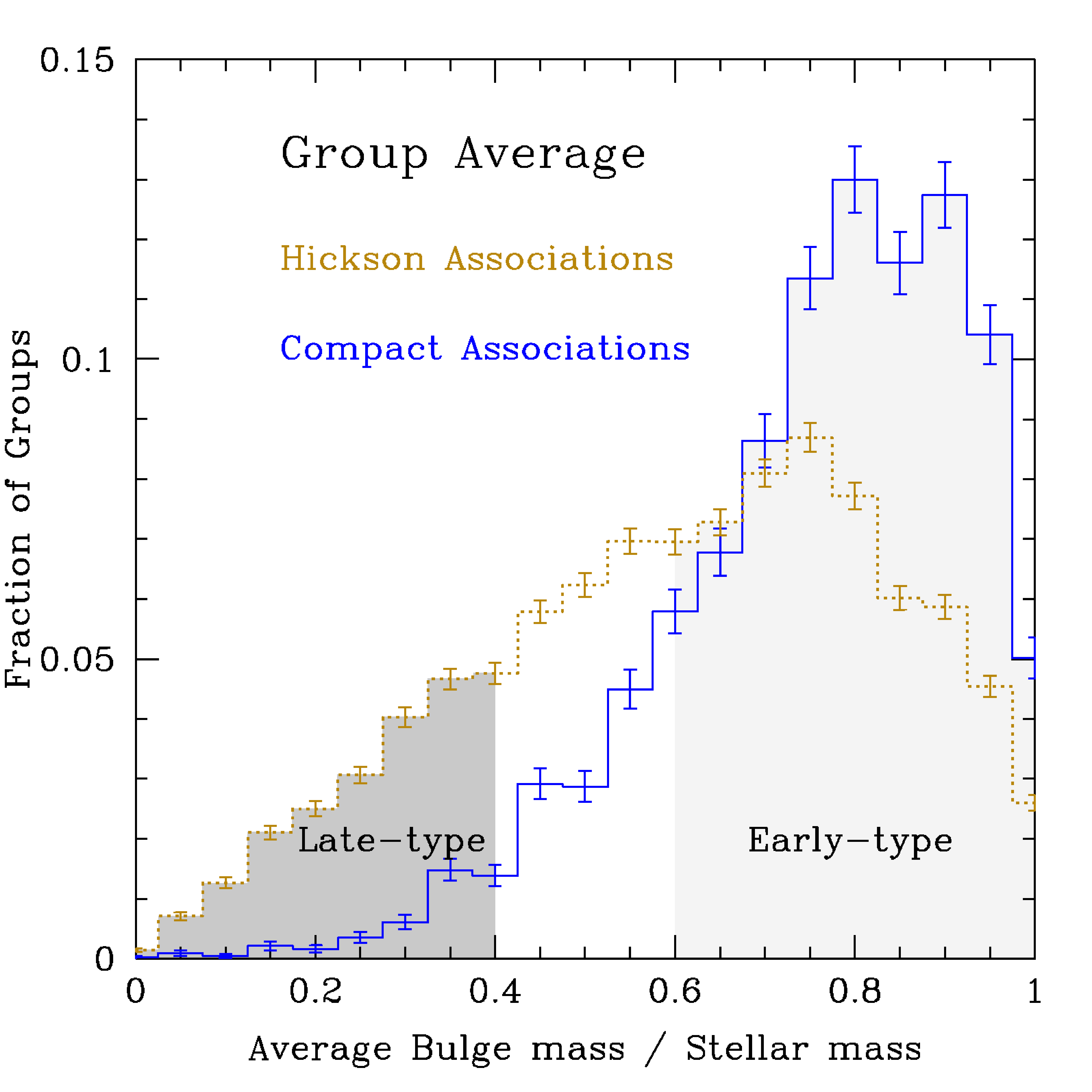}
    \caption{Left panel: the distribution of bulge-to-total stellar
      mass ($B/T$) for all galaxies in HAs, CAs and our control
      sample, with the same line-styles and scaling as
      Figure~\ref{stellarmassHCG}. Right panel: the average $B/T$ for
      each HA and CA, scaled to have the same total number of
      systems. The $B/T$ ratio can be used as a proxy for galaxy
      morphology, since late-type galaxies (spiral) have higher bulge
      fractions than early-type galaxies (elliptical). The shaded
      areas of the histograms show the approximate regimes of early
      ($B/T \gtrsim 0.6$) and late-type ($B/T \lesssim 0.4$)
      galaxies.}
    \label{B_T_HCG}
  \end{center}
\end{figure*}

The left panel of Figure~\ref{B_T_HCG} shows the distribution of $B/T$
for all the galaxies in the HAs, CAs and the control sample, where the
line-styles and normalization are the same as in
Figure~\ref{stellarmassHCG}.  The shaded parts of Figure \ref{B_T_HCG}
show the approximate regimes of late- and early- type galaxies. In the
left panel, there is a spike of galaxies in all three samples with
$B/T \sim 1$, ie. pure bulge galaxies. This is a result of the recipe
for bulge formation described in Section~\ref{Bulge_mass_fraction},
since a merger between two galaxies of approximately equal size leads
to most stars being deposited in a bulge. However, in this study we
are concerned only with the total fraction of early and late-type
galaxies. Regardless of whether these galaxies are really pure bulges
or not, it seems clear that they will be earlier type. In addition, all
three samples show the same feature, and so our relative results on
how galaxy morphology varies between the control, CAs and HAs will not
be affected.

The left panel of Figure~\ref{B_T_HCG} shows that galaxies in the HA
sample possess a much higher fraction of bulge-dominated galaxies than
the control sample, qualitatively consistent with the observational
results.  Using our definition for early-type galaxies, we find 55\,\%
of galaxies in HAs are bulge dominated, compared to only 35\,\% for
the control sample.

A cursory glance at the left panel of Figure~\ref{B_T_HCG} shows that
the effect of interloping galaxies is significant when examining the
relative proportions of early and late-type galaxies in CGs. In
particular, the $B/T$ distribution of HAs compared to CAs is
significantly different. Whereas the distribution for HAs has roughly
equal numbers of late- and early-type galaxies, the distribution of
CAs peaks at $B/T \sim 0.85$ and has a tail to smaller
values. Clearly, the preference for field galaxies to be late-type
means that interlopers artificially increase the proportion of
late-type galaxies which appear to be in CGs.  In total, 74$\%$ of CA
galaxies (ie. nearly three-quarters of all galaxies in CAs) are
bulge-dominated.

As well as examining individual galaxies, we can also determine the
average morphology of the galaxies in each HA and CA to see if they
are dominated by early or late-type galaxies. The right panel of
Figure~\ref{B_T_HCG} shows the average $B/T$ values for each HA and
CA. Only 4\,\% of CAs are late-type dominated groups (with an average
$B/T < 0.4$) whereas 84\% are dominated by early type galaxies (with
an average $B/T > 0.6$). This compares to $21$\% of late-type
dominated HAs and $54$\% of early-type dominated HAs.

Therefore we find a much higher fraction of early-type galaxies in CAs
compared with HAs, which tentatively suggests that the fraction of early-type
galaxies in observed CGs is likely to be under-estimated by $\sim$40\%
due to interlopers.

\subsection{Cold gas fraction and star formation rates}

\citet{WR87} obtained observations of neutral hydrogen (HI) in HCGs
and estimated the mass of cool gas within each group.  They found HCG
galaxies to be deficient in HI by a factor of two on average when
compared to galaxies with a similar morphology in loose groups.
\citet{HG1986} found that, in regions of higher galaxy density,
galaxies generally contain less HI compared to the field
population. This result was echoed by \citet{SG93}, and later
\citet{Maia1994}, who found that the neutral hydrogen content of
galaxies of the same morphology depended strongly on their
environment.  The low gaseous content of galaxies in high density
environments can be attributed to the removal of gas through processes
such as tidal and ram pressure stripping and/or enhanced star
formation activity during interactions, processes which may play an
important role in CG evolution.

\begin{figure}
  \begin{center}
    \includegraphics[width=84mm]{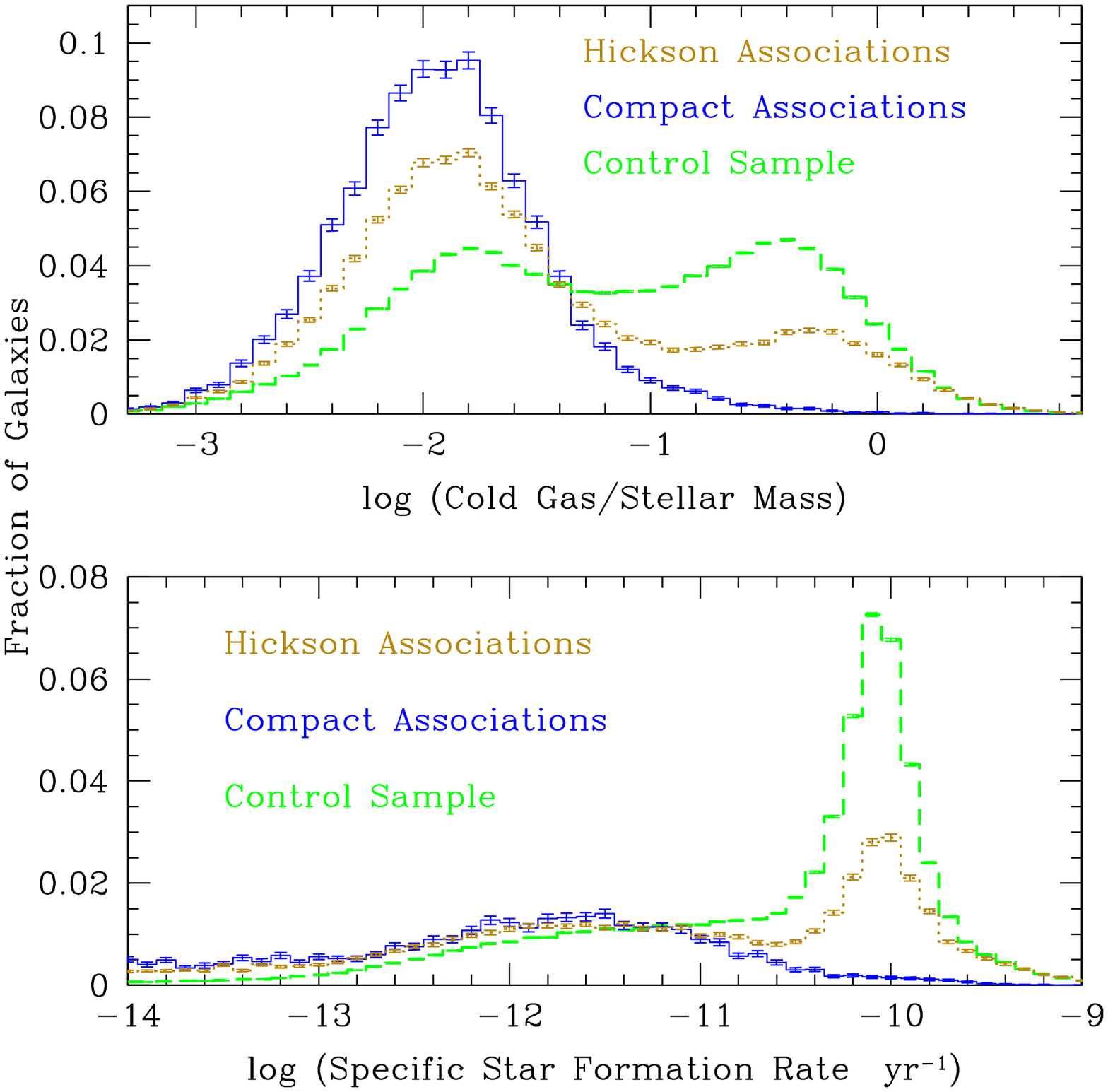}
    \caption{Top panel: Distribution of cold gas fractions (cold gas
      mass/total stellar mass) for galaxies in HAs, CAs and the
      control sample, with the same line-styles and scaling the same
      as in Figure~\ref{stellarmassHCG}.  Bottom panel: Specific star
      formation rates (star formation rate/stellar mass) above
      \mbox{$10^{-14}$\,yr$^{-1}$} for galaxies in HAs, CAs and the
      control sample, with line-styles and scaling as in
      Figure~\ref{stellarmassHCG}.  The control sample has a peak of
      relatively gas-rich, star forming galaxies (cold gas fraction
      $\gtrsim 10$\,\%, specific star formation rate $\gtrsim
      10^{-10.5}$\,yr$^{-1}$) and a large tail of gas poor galaxies
      with low specific star forming rates. Galaxies in HAs follow a
      similar distribution, but with a larger proportion of gas poor,
      low star-forming galaxies. In contrast, CAs have effectively no
      gas-rich, star forming galaxies.}
    \label{cold_sfr_HCG}
  \end{center}
\end{figure}

The upper panel of Figure~\ref{cold_sfr_HCG} shows the distribution of
the cold gas fraction of galaxies in HAs, CAs and the control sample,
expressed as the ratio of cold gas mass to stellar mass. The line
styles and scaling are the same as in Figure~\ref{stellarmassHCG}. The
control sample has a peak of relatively gas-rich galaxies with a
significant tail of gas-poor systems. Galaxies in HAs are
preferentially gas-poor relative to the control, although a
significant gas-rich population is also present: 23\,\% of galaxies in
HAs possess a cold gas fraction greater that 10\,\%, compared to
45\,\% of all control sample galaxies.

In contrast to both control and HA galaxies, CA galaxies
are nearly exclusively gas-deficient, with $<$4\,\% with
a cold gas fraction greater than 10\,\%. Once again, interlopers
appear to have the effect of increasing the apparent proportion of
relatively gas-rich galaxies in CGs. 

The bottom panel of Figure \ref{cold_sfr_HCG} shows the distribution
of specific SFR (SFR/stellar mass, hereafter SSFR) for all galaxies
with SSFR $> 10^{-14}$\,yr$^{-1}$.  The SSFR for HAs and the control
sample are quantitatively different, with the former having a
preference for lower values of the SSFR; 78\,\% of galaxies in HAs
have SSFRs lower than $10^{-11}$\,yr$^{-1}$, compared to 57\,\% for
the control. However, there are qualitative similarities between the
distributions, since both peak at $10^{-10}$\,yr$^{-1}$ and both have
a significant number of galaxies with much lower rates.

CAs have very low star formation rates on average, and a comparison of
the distribution of SSFR for galaxies in CAs compared to that for the
HAs is striking. Observations of HCGs suggest that they have SFRs
which are similar to field galaxies (e.g. \citealt{IP1999};
\citealt{S04}). However, our results suggest that this is once again
the result of interloping galaxies; CAs have effectively no galaxies
with SSFRs in excess of $10^{-10.5}$\,yr$^{-1}$.

\subsection{Group velocity dispersion}

\begin{figure}
  \begin{center}
    \includegraphics[width=84mm]{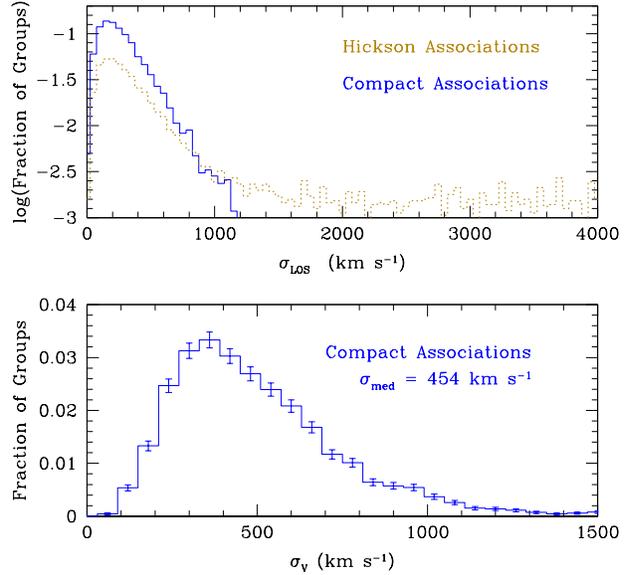}
    \caption{Top panel: the line-of-sight velocity dispersion
      ($\sigma_{LOS}$) distribution of HAs and CAs (dotted brown and
      solid blue line, respectively), where the histograms have been
      normalised to have the same total number of groups. Bottom
      panel: the three dimensional velocity dispersion of CAs. }
    \label{vel_disp}
  \end{center}
\end{figure}

One of the key observables for CGs, and one of the features which
makes them potentially such interesting systems to study, is their
line-of-sight velocity dispersion, $\sigma_{LOS}$. This is generally
measured to be relatively low ($\sigma_{LOS} \sim 200$\,km\,s$^{-1}$),
but could be influenced by the inclusion of interloping galaxies. To
see if this effect is significant, the top panel of
Figure~\ref{vel_disp} shows the distribution of line-of-sight velocity
dispersions for the HAs (brown dotted line) and CAs (solid blue line),
where the histograms have been scaled to have equal numbers of systems
in each. This velocity dispersion has been calculated from the
redshifts of each galaxy as listed in the mock catalogue, which
includes both the Hubble velocity and the peculiar velocity
components.

As the top panel of Figure~\ref{vel_disp} makes clear, the HAs have a
(very) significant tail to very high values of $\sigma_{LOS}$ due to
interloping galaxies at vastly different redshifts. Observationally,
any candidate member of a CG which was measured to possess a very
different redshift to the other members would be excluded from the
calculation of the velocity dispersion since that is a clear
indication that it is an interloper. For example, it is common
practice to exclude any group which has an apparent velocity
dispersion of $> 1000$\,km\,s$^{-1}$ (\citealt{HMH92}); in the
simulation, there are virtually no CAs with $\sigma_{LOS} >
1000$\,km\,s$^{-1}$, demonstrating the validity of this procedure.

The distribution of $\sigma_{LOS}$ below $1000$\,km\,s$^{-1}$ are very
similar for both CAs and HAs, with a peak at $\sim 150 -
200$\,km\,s$^{-1}$ and a mean value of $\sim 300$\,km\,s$^{-1}$ for
both samples. However, whereas there are $\sim 4200$ CAs in total,
there are $\sim 6600$ HAs with $\sigma_{LOS} <
1000$\,km\,s$^{-1}$. This implies that approximately one-third of all
CGs with $\sigma_{LOS} < 1000$\,km\,s$^{-1}$ consist, at least in
part, of interloping galaxies. These could act to contaminate any
further study of CG galaxy properties.

Finally, the bottom panel of Figure~\ref{vel_disp} shows the
three dimensional velocity dispersions of all the CAs identified in the
simulation. The peak de-projected velocity dispersion is at $\sigma_v
\sim 350$\,km\,s$^{-1}$, and the median velocity dispersion is
$\sigma_{med} \sim 450$\,km\,s$^{-1}$. CAs therefore are generally low
velocity dispersion systems, but it is interesting to note that there
is a tail to very high values and that some CAs have intrinsic
velocity dispersions in excess of 1000\,km\,s$^{-1}$.

\section{Observational comparison and discussion}
\label{Discussion}

In Section~3, we examined various (observable) properties of compact
groups found in a mock galaxy catalogue from \citet{LB07} with the aim
of comparing to observations of these systems. As with any such study,
uncertainties in the input physics of the simulation on which the mock
catalogue is based could affect our results. In this respect, our
results are as robust as modern galaxy simulations
allow. However, comparing a control sample to HA and CA galaxies
provides a very powerful way to test the effect interlopers have on
observed properties. Where possible, simulations are designed to
reproduce the results of more detailed numerical simulations (for
example in the case of galaxy mergers of specific mass ratios), and,
importantly, recreate many of the important observational properties
of low-redshift galaxy populations. In particular, it has been shown
that the luminosity function, the global star formation history, the
Tully-Fisher relation, the mass-metallicity relation, and the
colour-magnitude distribution are well matched to observations
(eg. \citealt{C06}), and the statistics of the clustering properties
of galaxies are also well reproduced (\citealt{S2005};
\citealt{Li2007}). \cite{M08} have recently used similar mock galaxy
catalogues to those used here to investigate the nature and properties
of close galaxy pairs and found good agreement with observations. Our
results should therefore provide a useful comparison to observational
samples of compact group galaxies.

Figure~\ref{colourmagdiagram} summarises some of our main results by
displaying the $\left(M_{u}-M_{r}\right)$ versus $M_r$
colour-magnitude distribution of galaxies in our samples. In both
panels, the green density map (with square-root scaling) shows the
distribution of galaxies in our field (control) sample. We
re-emphasise that our control sample consists of galaxies matched in
mass and redshift to HAs, and that these are drawn from the rest of
the mock catalogue, which has been shown to reproduce many of the key
properties of galaxies. In the left panel, the brown points show the
corresponding distribution of a random 30\,\% of galaxies identified
as belonging to HAs, and in the right panel the blue points show the
distribution for all galaxies in CAs. We now discuss this figure in
more detail in conjunction with the results presented in Section~3.

\subsection{The effect of interlopers}

\subsubsection{The colour-magnitude diagram}

\begin{figure*}
  \begin{center}
\includegraphics[width=86mm]{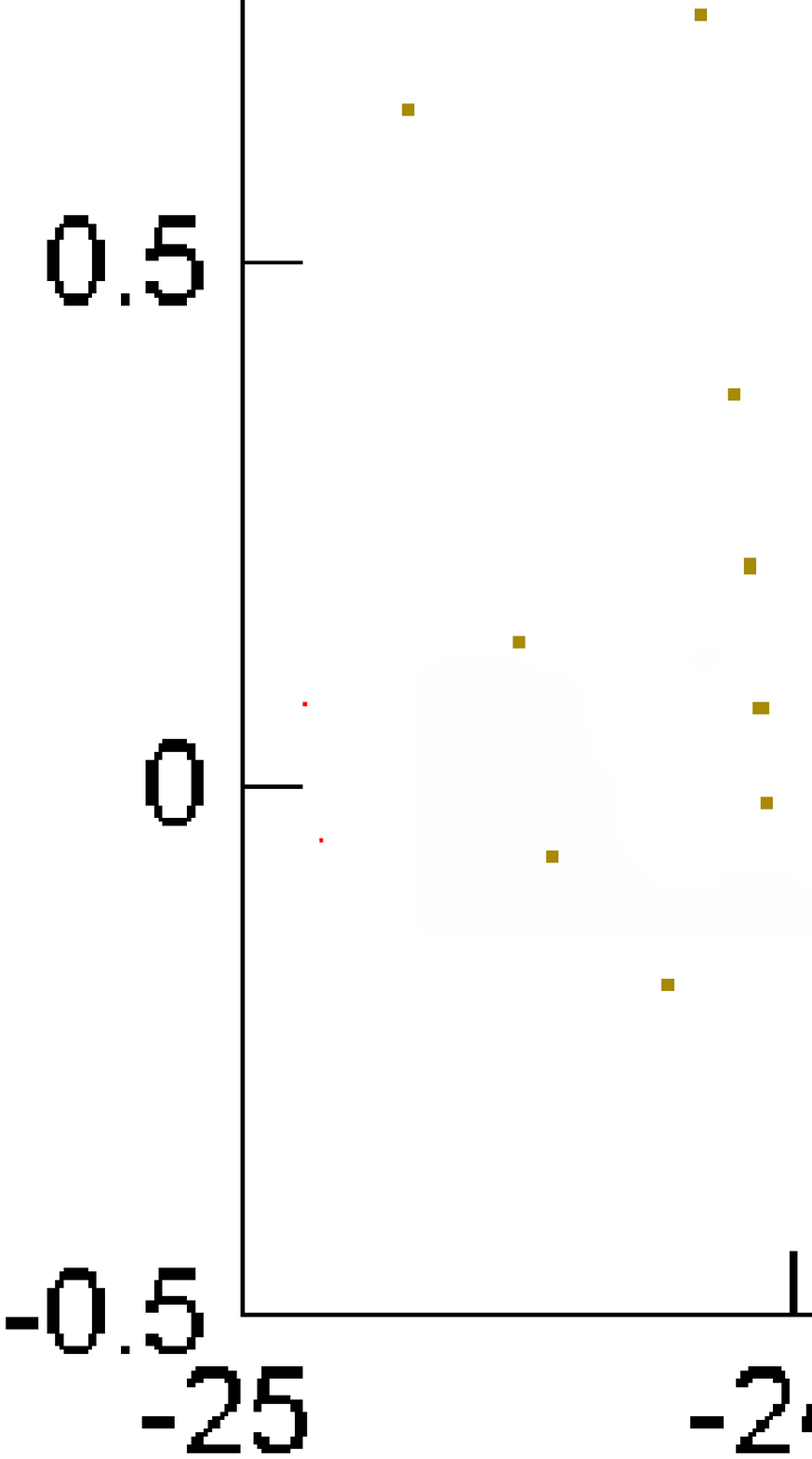}
\includegraphics[width=86mm]{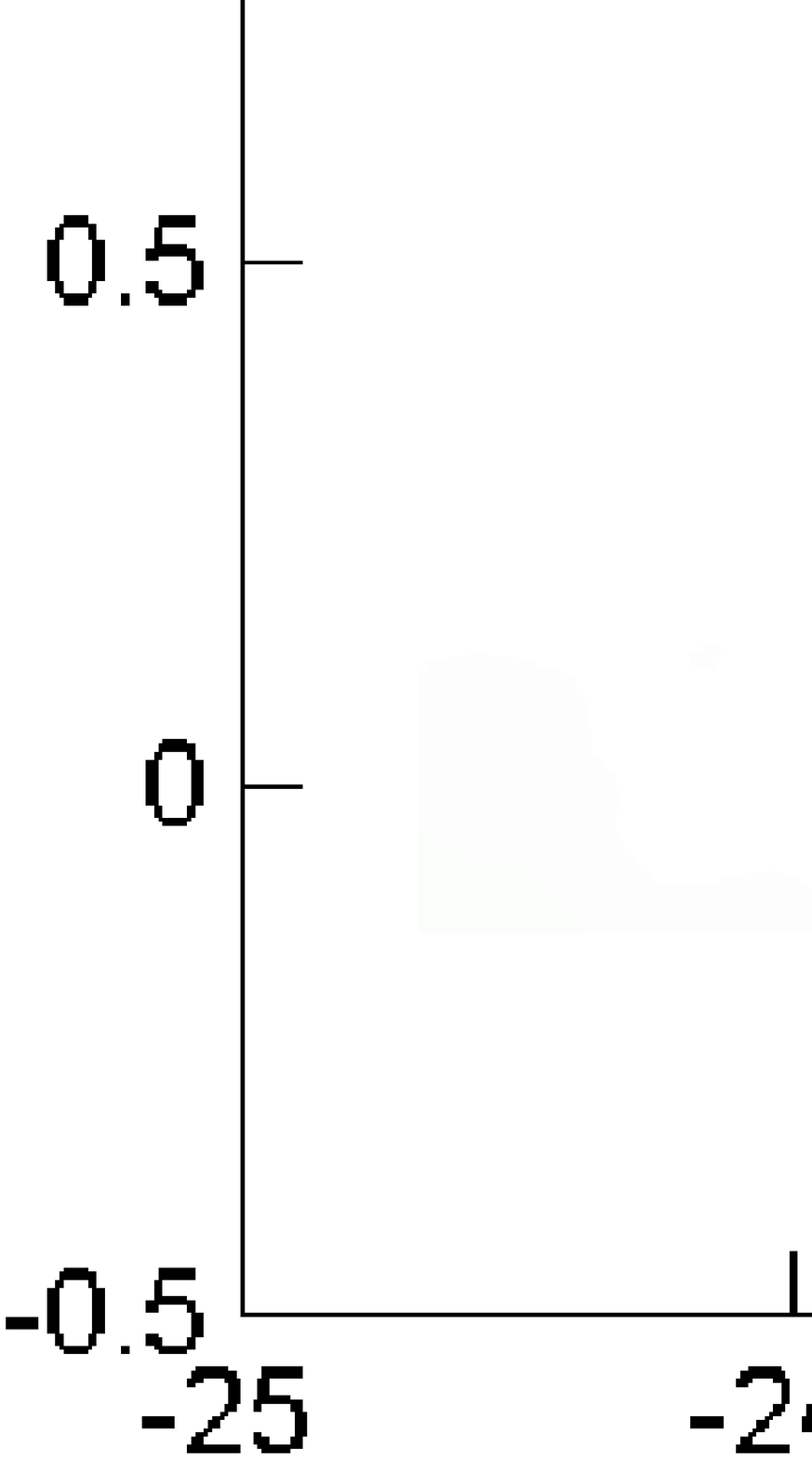}
 \caption{$M_r$ versus $\left(M_{u}-M_{r}\right)$ colour-magnitude diagrams
 for the galaxies in our sample. The green density map in both panels
 (with square-root scaling) is the distribution of the mass-matched
 control sample. The brown points in the left panel show the
 colour-magnitude distribution for a randomly selected 30$\%$ of
 galaxies in HAs, and the blue points in the right panel show the
 distribution for all the galaxies in CAs. The control sample shows
 the well studied red sequence and blue cloud populations. HAs show a
 prominent red sequence and have a significant number of bluer
 galaxies occupying the cloud. Most striking of all, the CAs
 effectively lack a blue population and nearly all galaxies are found
 in a very strong red sequence.}
\label{colourmagdiagram}
\end{center}
\end{figure*}

It is well known observationally that galaxies have a bi-modal colour
magnitude distribution (\citealt{St2001}; \citealt{Baldry2004}).  In
Figure~\ref{colourmagdiagram}, the control population shows this
distribution, with a relatively tight `red sequence' (centered around
$\left(M_{u}-M_{r}\right) \sim 2.7$) and a more diffuse `blue cloud'
(with $\left(M_{u}-M_{r}\right) \lesssim 2.3$) (\citealt{F05}). The
latter is populated predominantly by star-forming galaxies whereas the
former consists of galaxies with low star formation rates and older
stellar populations in general (\citealt{St2001,B2003,B2004,W2006}).

The left panel of Figure~\ref{colourmagdiagram} shows that the
galaxies in HAs (that is, those simulated galaxies identified on
application of the Hickson criteria) exhibit both a red sequence
population and a blue cloud population. In contrast, galaxies in CAs
(that is, those HAs which are physically dense and contain no
interlopers) are confined nearly exclusively to the red sequence, with
few galaxies occupying the blue cloud (97 $\%$ of CA galaxies are
redder than $\left(M_{u}-M_{r}\right) = 2.25$). The effect of
interlopers in compact groups therefore appears to introduce a
population of blue, star forming galaxies.

This conclusion is reinforced by our previous examination of colour
(Section~3.3), cold gas fraction and star formation rate (Section~3.4)
in these galaxies. We find galaxies in HAs are preferentially redder,
have lower gas fractions and have lower specific star formation rates
than the control sample, but in all cases bluer, higher gas fraction,
higher specific star formation rate galaxies are present in
significant numbers. Galaxies in CAs, however, are strikingly
different in all three observables. For example, 21\,\% of galaxies in
HAs are bluer than $\left(M_{u}-M_{r}\right) = 2.25$, 23\,\% of
galaxies in HAs have cold gas fraction in excess of 10\,\%, and 22\,\%
of galaxies in HAs have specific star formation rates higher than
$10^{-11}$\,yr$^{-1}$. In contrast, virtually no galaxies in CAs
satisfy any of these three criteria. In addition, $\sim 55\,\%$ of
galaxies in HAs appear to be bulge dominated, whereas the fraction of
bulge-dominated galaxies in CAs is significantly higher than this, at
$\sim 74\,\%$.

\begin{figure*}
  \begin{center}
    \includegraphics[width=60mm]{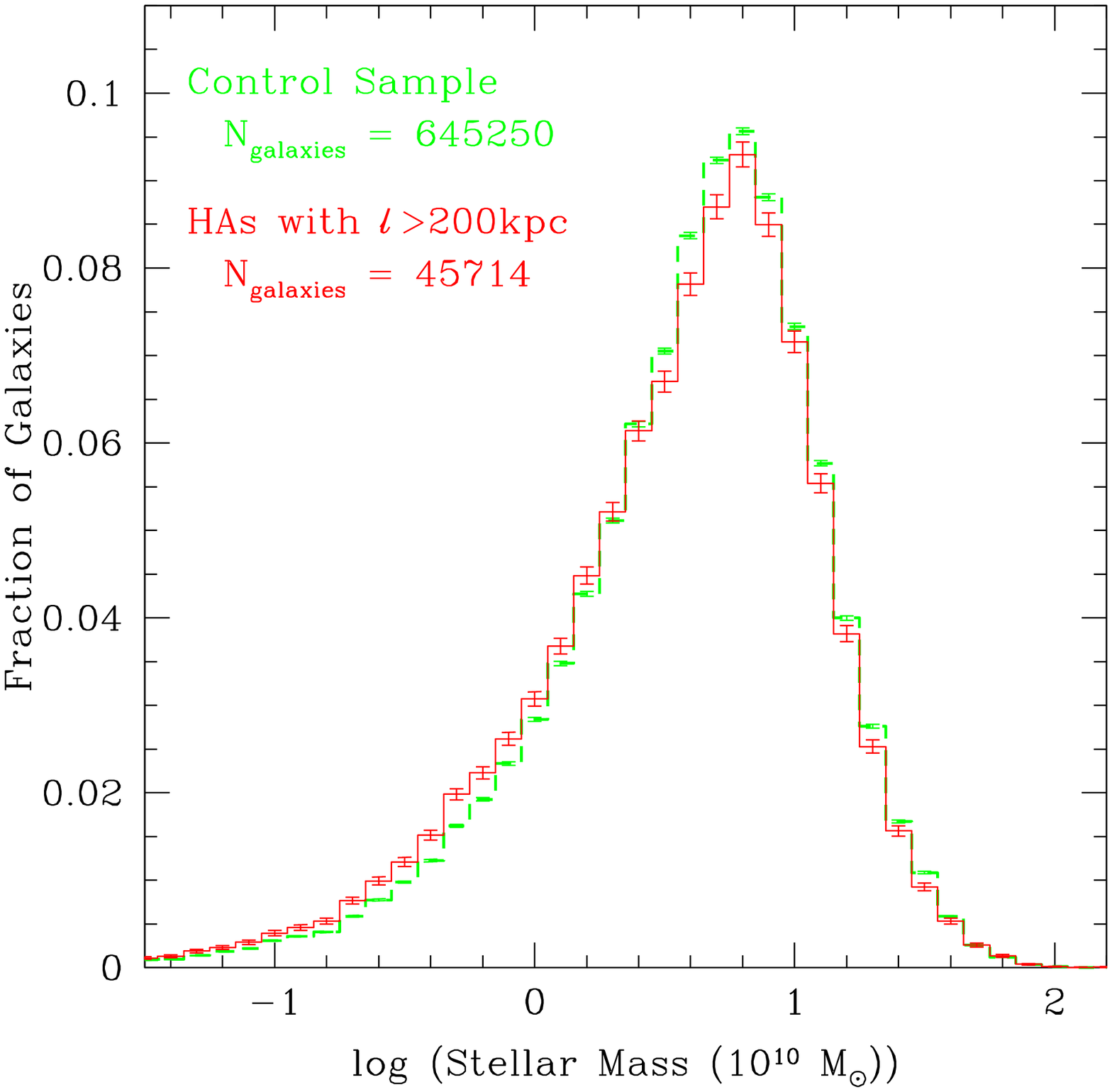}
 \includegraphics[width=60mm]{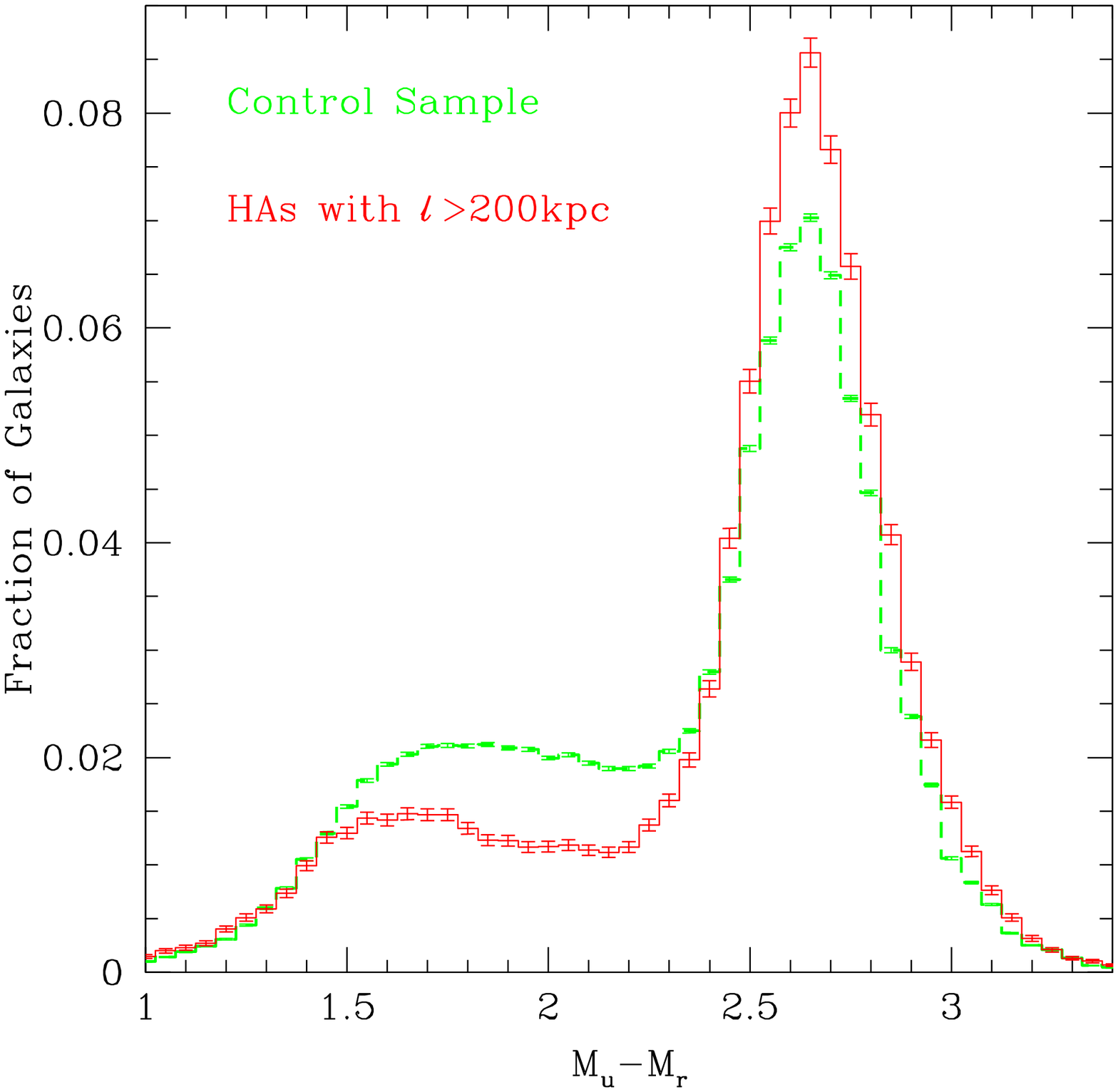}\\
    \includegraphics[width=60mm]{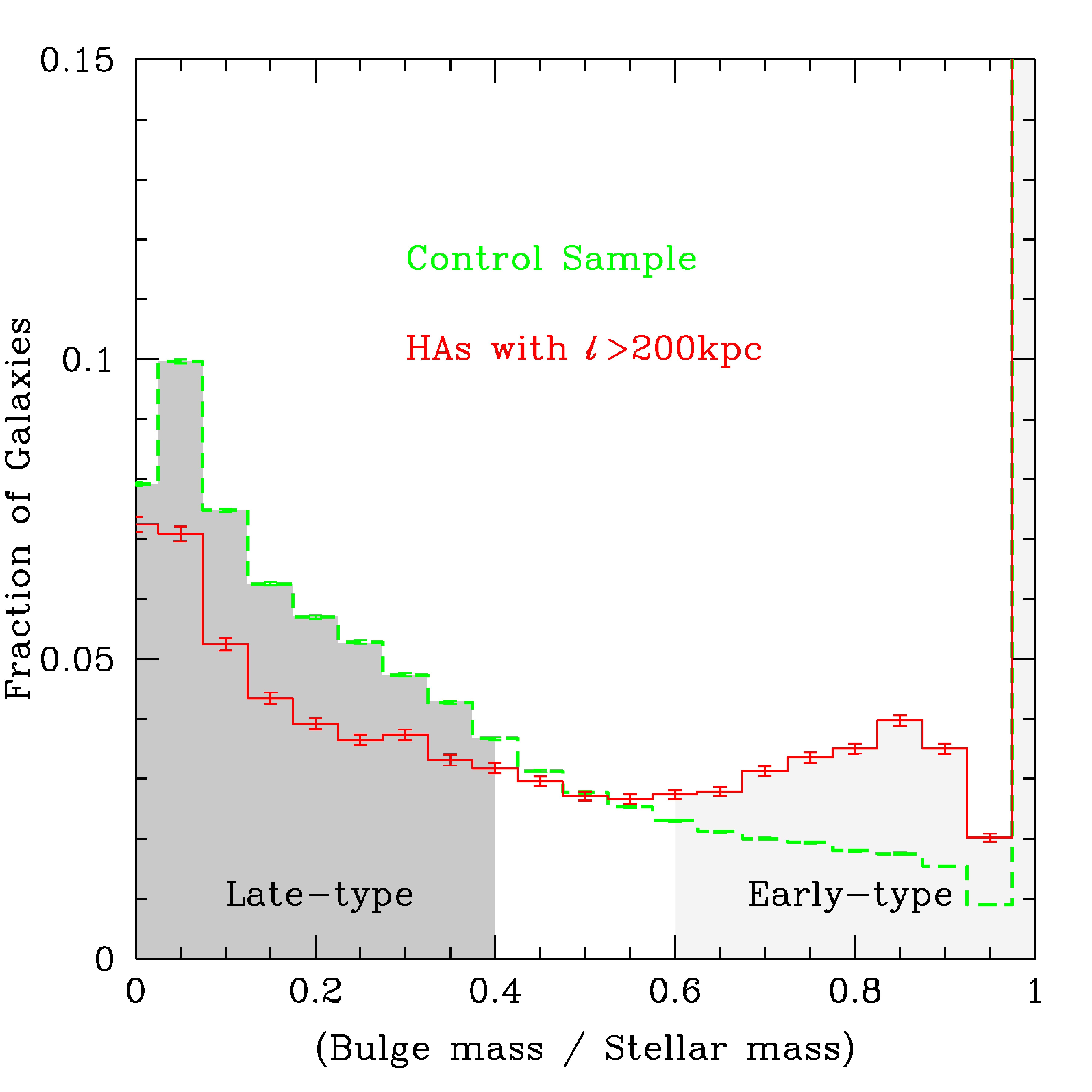}
    \includegraphics[width=60mm]{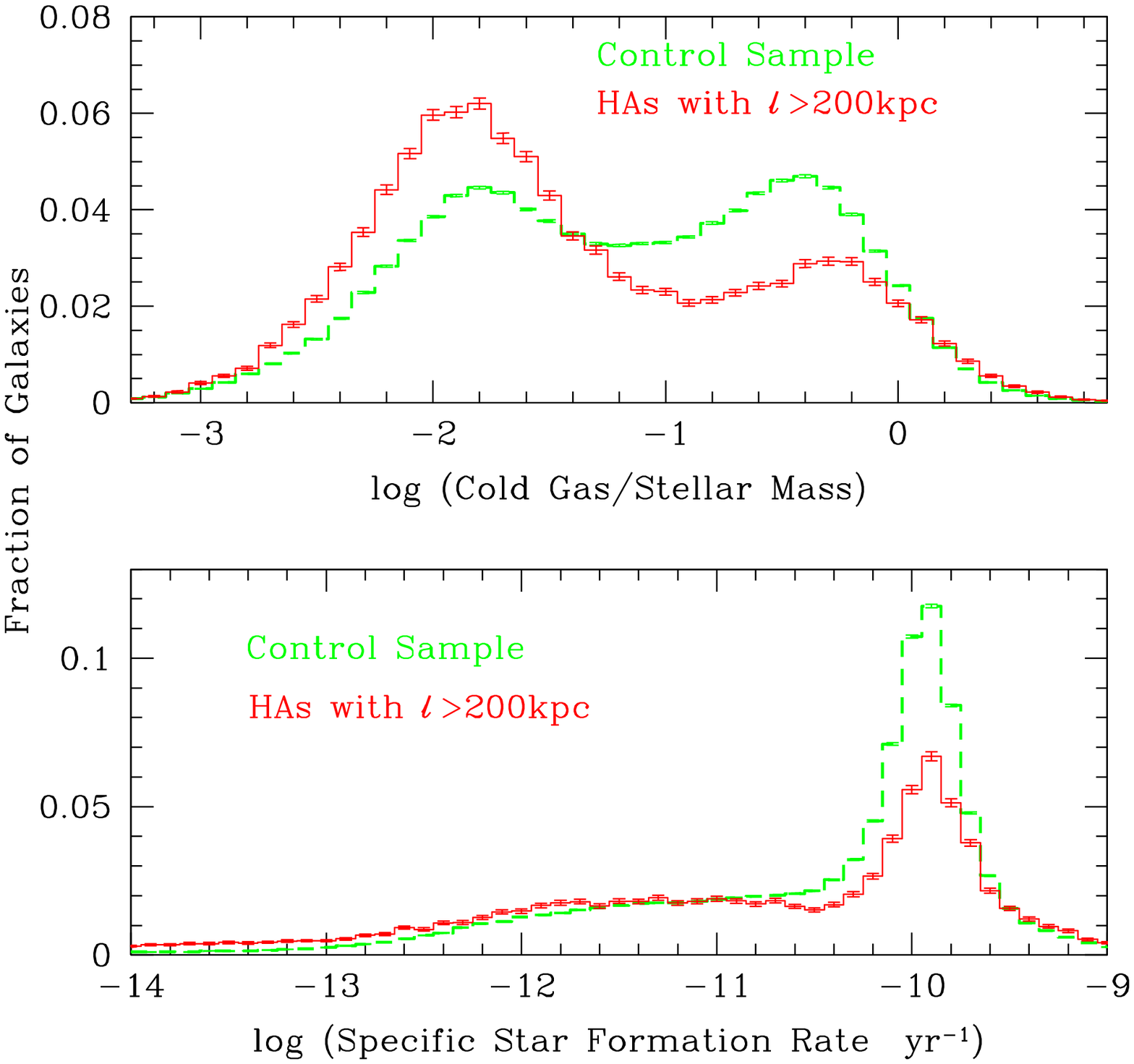}
    \caption{Distribution of various properties for the galaxies in
    the control sample (green dashed line) and for galaxies in the HAs
    which are {\it not} CAs (ie. galaxies in HAs with $\ell >$
    200~h$^{-1}$\,kpc and which have been classed as interlopers). The
    interlopers are clearly not randomly drawn from the field
    population but, like the control sample, there are a significant
    number of galaxies which are blue, star forming, gas-rich,
    late-type systems which act to bias our understanding of CGs.}
    \label{NotCAs}
  \end{center}
\end{figure*}

Figure~\ref{NotCAs} shows the distribution of the properties we have
been discussing for all galaxies in HAs which are {\it not} CAs
compared to the control sample. The galaxies in these
$\ell>$200~h$^{-1}$kpc HAs are clearly not randomly drawn from the
field, since their distributions differ from the control sample in all
panels of Figure~\ref{NotCAs}.  As we show in Paper I, these HAs
consist of some loose groups and sets of completely unassociated
galaxies, as well as many galaxy pairs, triplets, and even some
quadruplets, with unassociated background and/or foreground galaxies
projected along the same line-of-sight. It is this population of
interloping galaxies from the field which appears to bias our
understanding of the evolution of galaxies in CGs.

\subsubsection{Observational selection of compact groups}

We note that no cuts by velocity dispersion were applied when we
identified HAs. In practice, a velocity dispersion cut can only be
applied when a relatively small number of compact groups are being
studied, since getting redshift information for every galaxy in every
group of a much larger sample is impractical. For example, in the
SDSS, galaxies are quasi-randomly targeted for spectroscopy and the
probability that all galaxies in a compact group are selected for
spectroscopy is very small, especially since fiber collisions will
occur if trying to target all the members of a dense system. In the
few situations where full redshift information is available, groups
with very large velocity dispersions ($\sigma_{LOS} \gtrsim
1000$\,km\,s$^{-1}$) are usually discarded (eg. \citealt{HMH92}) since
it is likely they contain interlopers. Our analysis in Section~3.5
suggests that such an approach can greatly reduce contamination;
however, over one-third of the systems identified are predicted to
still contain interlopers and are therefore not genuine compact
groups. This population will still potentially act to bias any study
of CG properties.

As discussed in Section~4.1.1, $97$\,\% of galaxies in CAs are redder
than $\left(M_{u}-M_{r}\right) = 2.25$. Our analysis of the mock
catalogue therefore suggests that the purity of a CG sample can be
significantly increased by only selecting those groups whose members
are redder than $\left(M_{u}-M_{r}\right) \simeq 2.25$, since this
will preferentially remove HAs which have $\ell >$ 200~h$^{-1}$
kpc. Of course, selecting CGs by colour may lead to a potential bias,
since we select against any genuine CGs which happen to have bluer
galaxies. However, for studies of CGs where the property of interest
is not thought to depend strongly on colour, then this selection
technique potentially offers the opportunity to greatly reduce
contamination. Projects such as the Red Cluster Survey
(\citealt{GY2005}) adopt a similar methodology for the identification
of galaxy clusters.

\subsection{Observational comparisons}

There is an extensive body of literature on the observational
properties of galaxies in CGs and, in general, we find good
correspondence between the properties of galaxies identified in HAs in
the mock catalogue and those in reality.  Observationally, CGs contain
a greater proportion of early-type galaxies than the field;
\citet{H88} and \citet{P95} found that roughly half of all galaxies in
CGs were early type, compared to $\sim 20\,\%$ for the field
(\citealt{N73,G80}). We find that $\sim 55\,\%$ of all galaxies in HAs
in the mock catalogue (ie. those galaxies identified as belonging to
compact groups through application of the Hickson criteria) have $B/T
\ge 0.6$, and are probably best described as early-type. As discussed
in the previous section, however, the true fraction of early type
galaxies in compact groups is predicted to be nearer three-quarters,
since interlopers from the field (which are more likely to be
late-type) can increase the apparent proportion of late-type galaxies
in CGs.

\citet{L04} and \citet{D07} both find that galaxies in compact groups
are redder than the field population on average. Our results suggest
the same; 79\,\% of galaxies in HAs are redder than $\left(M_u -
M_r\right) = 2.25$, compared to only 63\,\% for the field. \citet{L04}
measure median colours for their compact group galaxies of
$\left(M_{r*} - M_{i*}\right)_{med} = 0.39$ and $\left(M_{u*} -
M_{g*}\right)_{med} = 1.66$. Our values for the galaxies in HAs
compare very favourably to this, with $\left(M_r - M_i\right)_{med} =
0.39$ and $\left(M_u - M_g\right)_{med} = 1.66$. For the field
samples, \cite{L04} calculate $\left(M_{r*} - M_{i*}\right)_{med} =
0.38$ and $\left(M_{u*} - M_{g*}\right)_{med} = 1.58$, whereas we find
$\left(M_{r*} - M_{i*}\right)_{med} = 0.38$ and $\left(M_{u*} -
M_{g*}\right)_{med} = 1.61$.

\begin{figure}
  \begin{center}
    \includegraphics[width=84mm]{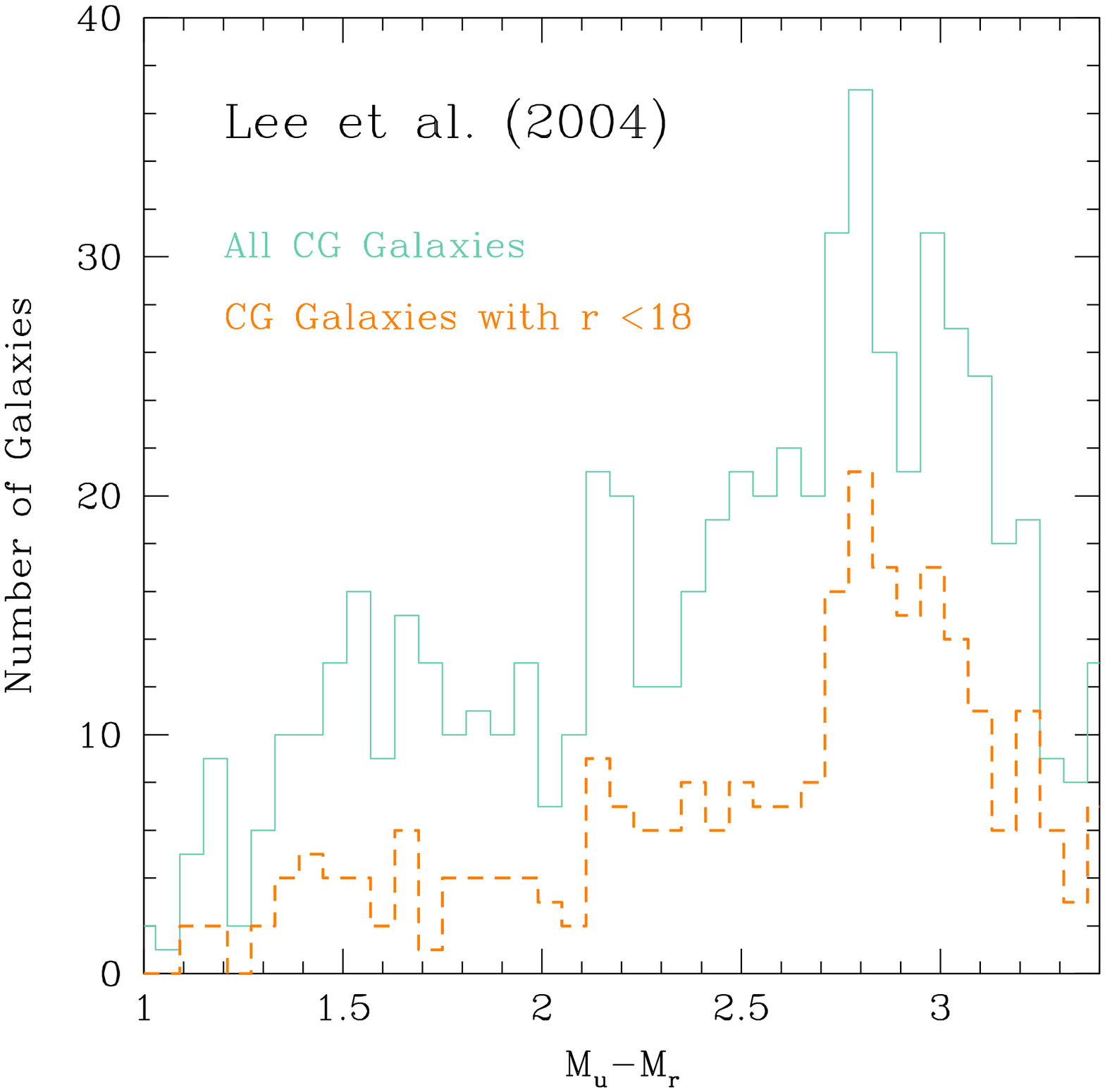}
     \caption{The observed $\left(M_u - M_r\right)$ colour
     distribution for compact group galaxies identified by \citet{L04}
     corrected for foreground extinction. Compact groups in the
     \citet{L04} catalogue were identified using a galaxy catalogue
     with a magnitude limit of $r = 21$, three magnitudes deeper than
     our mock galaxy catalogue. The blue histogram shows all 744
     galaxies in the observed CGs, whereas the orange histogram shows
     only those galaxies brighter than $r = 18$, the magnitude limit
     of our mock catalogue (although note that this subset of observed
     galaxies is different to those galaxies which would be identified
     had the original catalogue been magnitude limited at $r = 18$).}
    \label{colourlee}
  \end{center}
\end{figure}

Figure~\ref{colourlee} shows the observed (de-reddened) $\left(M_u -
M_r\right)$ colour distribution for the \citet{L04} compact
groups. The \citet{L04} catalogue identifies compact groups using a
galaxy catalogue with a magnitude limit of $r = 21$, three magnitudes
deeper than our mock galaxy catalogue. The blue histogram therefore
shows all 744 galaxies in the observed CGs, whereas the orange
histogram shows only those galaxies brighter than $r = 18$, the
magnitude limit of our mock catalogue (although note that this subset
of observed galaxies is different to those galaxies which would be
identified had the original catalogue had a magnitude limit of $r =
18$).

There are several qualitative and quantitative similarities between
the observed colour distributions in Figure~8 and the predicted ones
in Figure~2. In particular, there is a red peak with a blue tail in
the data, with the peak at $\left(M_u - M_r\right) \sim 2.7$, similar
to Figure~2. 68\,\% of CG galaxies with $r \le 21.0$ are redder than
$\left(M_u - M_r\right) = 2.25$ (75\,\% of CG galaxies with $r \le
18.0$). This compares favourably to the 79\% in the same colour range
from the simulation, although it may suggest that the simulated
galaxies are slightly too red compared to observations. However, the
reddest galaxies are seen in the observations; whereas there are very
few simulated CG galaxies with $\left(M_u - M_r\right) > 3$, 30\,\% of
the galaxies in Figure~8 are redder than this limit.

\citet{WR87} found that galaxies in compact groups appear deficient in
HI by a factor of approximately two in comparison to the field. While
it is not possible to calculate the fraction of HI in the galaxies in
the simulation, it is possible to determine the fraction of `cold gas'
in galaxies in HAs and CAs relative to the field. We have made the
assumption that the cold gas fraction traces the HI gas fraction and
the fraction of other atomic or molecular gases should be
negligible. Once again, we find that galaxies in HAs have, in general,
a considerably lower cold gas fraction than the control sample (77\,\%
of galaxies in HAs have a cold gas fraction less than 10\,\%, compared
to 55\,\% for the control). The mean cold gas fraction of the control
sample galaxies is $\sim 6 - 7$\,\%, whereas the mean cold gas
fraction for galaxies in HAs is approximately a factor of two smaller,
at $\sim 3\,\%$. This is in good relative agreement with observations,
although we note that our control sample is not made up of galaxies
exclusively in loose groups, as was the case for the observational
study.

The main disagreement between our study and observational work is the
comparison of SSFRs.  Observational studies suggest that the star
formation rates of galaxies in CGs are broadly similar to the general
field population (\citealt{IP1999}). Inspection of the lower
panel of Figure~\ref{cold_sfr_HCG} shows that the SSFRs of field
galaxies and HAs are quantitatively different, but they do share some
similarities. In particular, they both have peaks at $\sim
10^{10}$\,yr$^{-1}$ and tails to much lower values. 
It is possible that the observations do not yet have sufficient statistics to show the differences in the distributions, although it is equally possible that the differences do not exist.
It is once again important to emphasise, however,
that any similarities between the star formation rates of HAs and the
control sample disappear when one considers only
those groups which are physically dense and which contain no
interlopers.

These trends are also seen for galaxies in cluster environments:
\citet{GH83} found spiral galaxies in the Virgo cluster were deficient
in HI relative to their counterparts in the field. Lower star
formation rates for galaxies in clusters compared with those in the
field at the same redshift have also been found up to z $\sim$ 1
(\citealt{K83}; \citealt{Balogh1997}; \citealt{Hashimoto1998};
\citealt{Poggianti1999}; \citealt{Postman2001};
\citealt{Gomez2003}). \citet{BO84} found that compact clusters at low
redshift had cores which were essentially devoid of blue galaxies and
a clear correlation between galaxy morphology with local density
(\citealt{D80}; \citealt{D97}) and cluster-centric radius
(\citealt{W93}) has been observed.

It is reassuring to find that the observational properties of CGs are
broadly reflected in the properties of the galaxies identified in the
mock catalogue; these galaxies are generally more likely to be
bulge-dominated systems than the field, are redder, have lower cold
gas fractions, and exhibit relatively similar star formation rates to
the field population. In colour-magnitude space, galaxies identified
as belonging to compact groups (HAs) occupy a dominant red sequence
but there are considerable numbers which occupy the blue cloud. On the
removal of interlopers, nearly all galaxies in compact groups occupy
the red sequence (Figure~\ref{colourmagdiagram}).

\subsection{On the homogeneity of compact associations}

Our results agree well with the observations of CGs, but in addition
strongly suggest that some of these results have been influenced by
interloping galaxies which did not co-evolve with the other galaxies
in the CG. When this is taken into account, the simulations suggest
that the fraction of galaxies in CGs which are {\it bulge-dominated,
red, have low cold gas fractions and low star formation rated have
been previously underestimated}. The low star formation rates that we
imply are consistent with evolution in an environment where
interactions were common, where either most of the gas was initially
removed from these galaxies via stripping processes and/or the gas was
used up in brief, intense periods of star formation shortly after the
galaxy entered the CG environment.

What is very noticeable from our results, and is nicely summarised in
the right panel of Figure~\ref{colourmagdiagram}, is the homogeneity
of the galaxies in CAs; virtually all belong to the red sequence and
are ``red and dead''. We first discuss if the origin of this result is
due to a simplified treatment of gas removal in the semi-analytical
modeling of satellites in these simulations (Section 4.3.1). On
concluding that this is not the likely interpretation, we turn our
attention to possible evolutionary scenarios for compact groups
(Section 4.3.2).

\subsubsection{Central and satellite galaxy evolution}

There are certain evolutionary processes which are believed to operate
predominantly on satellite galaxies, and which do not affect central
galaxies (that is, those at the centres of their own dark matter halo)
to the same degree. In particular, when a smaller dark matter halo is
accreted by, and becomes a satellite of, a larger dark matter halo,
its hot, diffuse gas may be stripped. This gas can no longer feed the
satellite galaxy, and so once the cold gas in the galaxy's disc is
used up, its star-formation ends (termed ``strangulation'';
\citealt{LTC80}; \citealt{Balogh2000}). In circumstances where the
external pressure is sufficiently high, ram pressure stripping may
remove the cold gas directly, abruptly ending star formation
(\citealt{GG72}).

In most semi-analytic models, the loss of a satellite's hot gas is
modeled as an instantaneous process. However, it has been suggested
that this not a good approximation in all cases
(e.g. \citealt{McC2008}; \citealt{DB06}) and it has been shown that
satellites simulated in this way are frequently too ``red and dead''
in comparison to observational results
(\citealt{Wein2006,Baldry2006,GB2008,Kang2008}). In Paper~I, we showed
that just over half of CAs in our mock catalogue share a common dark
matter halo, where one galaxy is considered to be the ``central''
galaxy and the remaining ones are considered ``satellites'' (although
CAs consisting of 4 separate haloes, and thus four ``central''
galaxies, do exist). Given the significant number of ``satellite''
galaxies in our sample, it is important to determine if uncertainties
in the treatment of strangulation for these objects has lead us to
conclude, incorrectly, that CAs are predominantly red and dead.

\begin{figure*}
  \begin{center}
    \includegraphics[width=86mm]{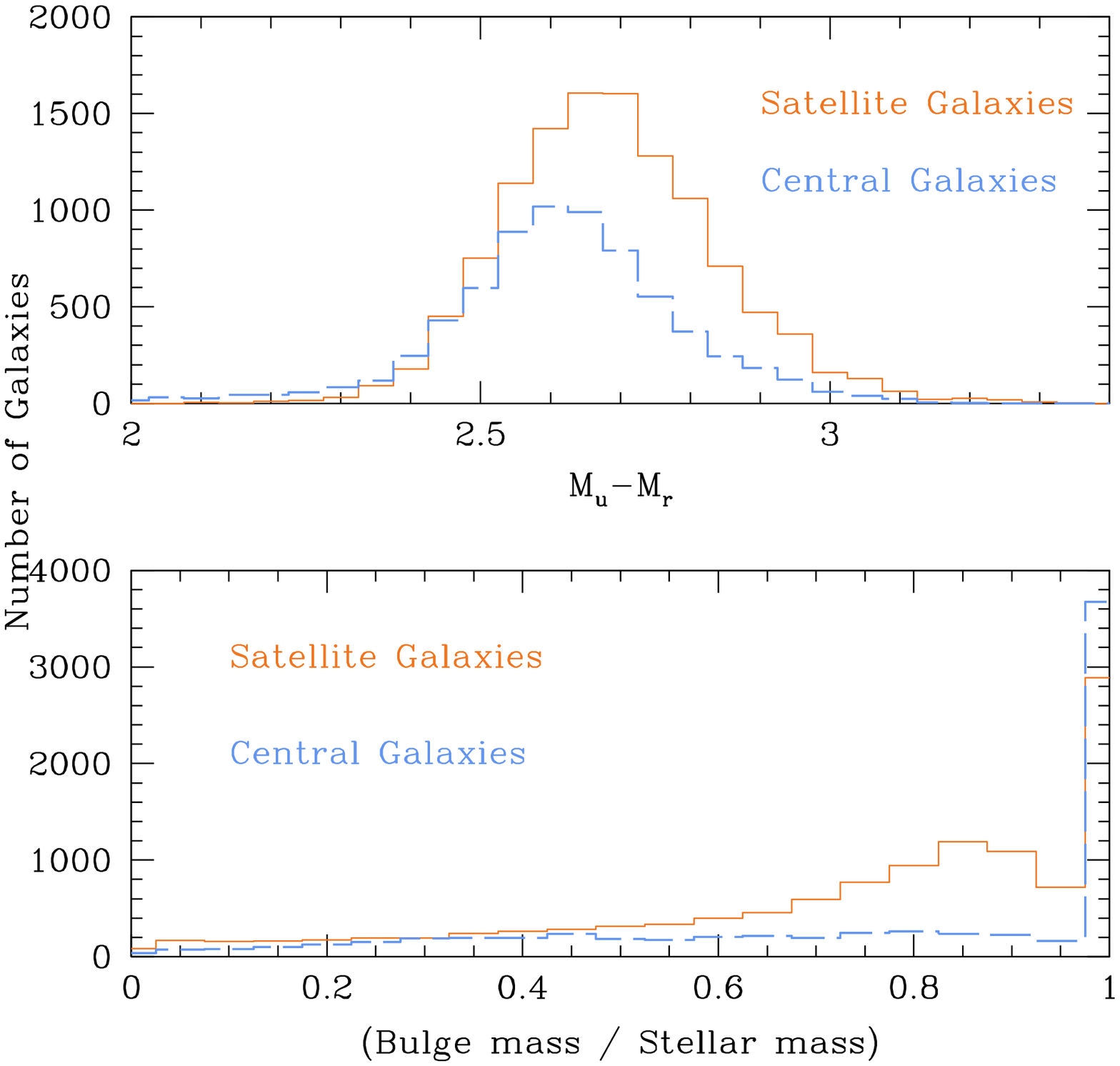}
    \includegraphics[width=86mm]{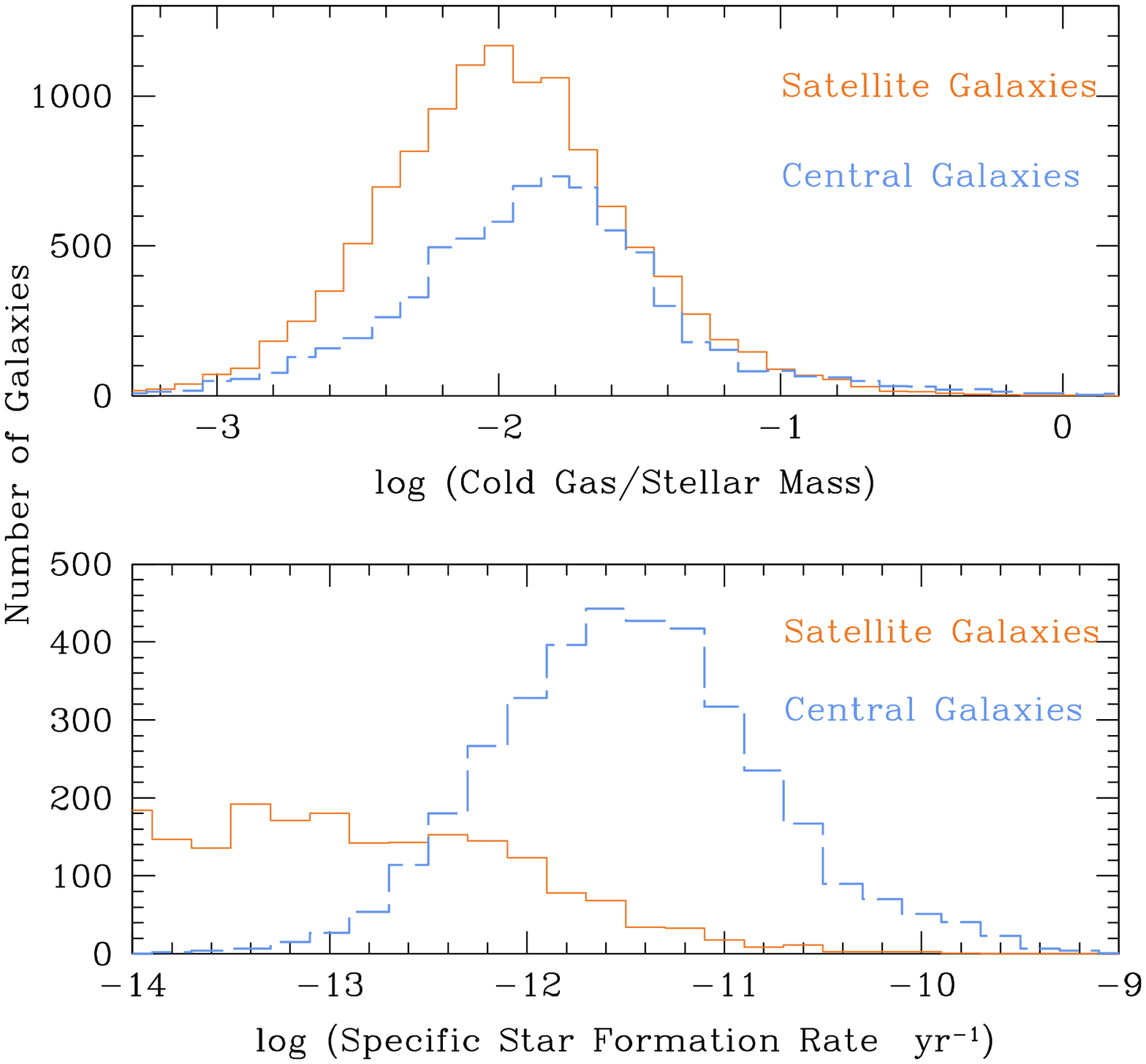}
    \caption{Shown above are the distributions for colour (upper
      left), cold gas fraction (upper right), $B/T$ (lower left), and
      SSFR (lower right) for all galaxies in CAs divided into two
      sub-sets; ``satellite'' galaxies (orange solid) and ``central''
      galaxies (dashed blue) (11634 and 7177 galaxies in each sub-set,
      respectively). }
    \label{Haloproperty}
  \end{center}
\end{figure*}

Figure~\ref{Haloproperty} shows the colour (top left), cold gas fraction (top right),
$B/T$ (lower left) and SSFR (lower right) for all the galaxies which
we identify as being a member of a CA in our mock catalogue, split
according to whether they are a ``satellite'' (orange solid lines) or
a ``central'' galaxy (blue dashed). Central galaxies of CAs are
marginally bluer than satellites and have a slightly higher cold gas
fraction. The central galaxy SSFR is significantly higher than for the
satellites, reflecting the fact that strangulation has quenched star
formation in satellites. The morphology of central galaxies in CAs, as
represented by $B/T$ (see Section 3.3), is a mix of early- and
late-type but with a dominant peak around $B/T = 1$, which indicates
that a major merger has occurred (recall Section 2.2.4). There is a
similar peak for satellite galaxies, which show a preference towards
having a higher bulge fraction than central galaxies.

From Figure~\ref{Haloproperty}, we conclude that ``central'' galaxies in CAs show some
differences to ``satellite'' galaxies in CAs, particularly regarding
the SSFR. However, it is still the case that, for both central and
satellite galaxies in CAs, they are much redder, have a considerably
smaller cold gas fraction, are generally earlier type (as traced by
bulge fraction) and have significantly lower SSFR than the field
galaxies in our sample (recall Figures 2, 3, and 4). Thus our overall
conclusion about CAs being ``red and dead'' galaxies, significantly
redder than the field population, stands even if we only consider
``central'' galaxies. A simplified treatment of satellite
strangulation cannot account for this result.

In the semi-analytic models, central galaxies are typically red if they are massive (where AGN feedback can offset a slowly cooling halo) whereas satellite galaxies tend to be red, independent of their stellar mass (\citealt{Wein2006}).
To ensure that the high red fraction of CAs is not driven by a large (red) satellite population and massive (red) central galaxies, we examined the colour distribution of central CA galaxies as a function of stellar mass. At the low mass end, central galaxies in CAs still dominate the red sequence whereas galaxies in the control sample are still mostly in the blue cloud.

Finally, we note that central galaxies in the HAs, CAs and the control sample
(matched in mass and redshift to the HAs) are all subject to the same
uncertainties in the semi-analytic modeling (e.g., quenching, AGN
feedback). Thus, the {\it relative} distribution of galaxy properties
in each of these three subsets provide a more robust comparison than
is possible using absolute values.

\subsubsection{The evolution of compact associations}

Our current comparison between compact groups identified in a mock catalogue
to previous observations of compact groups suggests that the former
reproduce the latter well, given the current theoretical and
observational limitations. However, the fact that the simulated CAs do
not resemble typical field galaxies, and in fact show only a small
spread in properties, leads us to consider possible evolutionary
explanations which could produce this homogeneous population:

\begin{enumerate}
\item most compact groups did not form out of the typical field galaxy
population, and instead formed out of a population of ``red and
dead'', early-type galaxies;

\item most compact groups formed long ago, perhaps out of the typical
field galaxy population, and the member galaxies evolved over
cosmological timescales to become a population of ``red and dead''
ellipticals;

\item compact groups form continuously, perhaps out of the typical
field galaxy population, and the evolutionary times of galaxies in
compact groups are very short so they quickly become a population of
``red and dead'' ellipticals.
\end{enumerate}

The origin for the different properties of CA galaxies compared to the field in these models must be rooted in
their mass assembly and gas accretion histories.  
In a future contribution, we will explore the dynamical evolution of
the simulated compact groups in the semi-analytic simulations by
tracing the progenitors of the galaxies and haloes identified as
belonging to a CA in our $z = 0$ mock catalogue using the outputs of
the simulation from previous time steps. This enables us to reconstruct
the formation history of the groups and compare their lifetimes to the
evolutionary timescales of galaxies, particularly those relating to
movement onto the red sequence of the colour magnitude diagram.

\section{Summary}

In this paper we have compared the observed morphological properties
of galaxies in compact groups with those identified in a mock galaxy
catalogue, to determine if and where our observational understanding
of the evolution of these systems is in conflict with modern
theoretical simulations of galaxy evolution. We find that the
properties of galaxies identified as belonging to compact groups in
the simulation are in qualitative agreement with observations. Bluer
galaxies, late-type galaxies, gas-rich galaxies and galaxies with
higher specific star formation rates are all present in compact
groups.  However, on average, compact groups exhibit a higher
proportion of early-type galaxies than the field, and are generally
redder and gas-deficient compared to the field. The specific star
formation rates of compact group galaxies are, on the other hand,
relatively similar to the field.

Importantly, we find that the effect of interlopers on these results
is significant. By using the full three dimensional positions of the
galaxies in the mock catalogue, we can remove those groups which are
not physically dense and/or which contain interlopers. The compact
groups which remain are found to consist nearly exclusively of red
$\left(M_{u}-M_{r}\right) \gtrsim 2.5$, gas-deficient ($M_{cold\,gas}
< 0.1 M_{*}$) galaxies with low specific star formation rates (SSFR $<
10^{-10.5} yr^{-1}$). 

We predict that approximately $84\,\%$ of compact groups are dominated
by early-type galaxies when interlopers are accounted for, compared to
$\sim 54\,\%$ when they are not. Thus modern cosmological simulations
suggest that galaxies in compact groups have homogeneous properties
and are predominately ``red and dead''. In contrast, observations
suggest that compact groups exhibit a larger range in galaxy
properties. Our results imply that the majority (but not all) of
galaxies identified as being members of compact groups which are
late-type, blue, relatively gas rich and/or have relatively-high star
formation rates are interlopers. Selection by colour is predicted to
greatly reduce contamination levels of CGs, although this will
introduce a potential bias into the population. Where redshift
information is available for all prospective group members,
observational contamination can be reduced; however, one-third of all
compact groups with low velocity dispersions ($\sigma_{LOS} <
1000$\,km\,s$^{-1}$) appear to contain interlopers.

\section*{Acknowledgments}
The Millennium Simulation databases used in this paper and the web
application providing on-line access to them were constructed as part
of the activities of the German Astrophysical Virtual Observatory. We
thank Luc Simard as well as Chien Peng for useful discussions relating to this work, and the
anonymous referee for several useful suggestions which improved this
paper. AWM acknowledges support from a Research Fellowship from the
Royal Commission for the Exhibition of 1851. He also thanks Sara
Ellison and Julio Navarro for additional financial assistance.  SLE
and DRP acknowledge the receipt of NSERC Discovery Grants which funded
some of this research.


\begin{thebibliography}{}

\bibitem[Alonso et al. (2004)]{Alonso2004}Alonso, M. S., Tissera, P. B., Coldwell, G., \& Lambas, D. G. 2004, MNRAS, 352, 1081

\bibitem[Andernach \& Coziol (2007)]{AC07} Andernach, H., \& Coziol, R. 2007, in Groups of Galaxies in the Nearby Universe, ed. I. Saviane, V. Ivanov, \& J. Borissova (Berlin: Springer), in press (astro-ph/0603295)

\bibitem[Baldry et al. (2004)]{Baldry2004} Baldry, I. K., Glazebrook, K., Brinkmann, J., Ivezi\'c, \^Z., Lupton, R. H., Nichol, R. C., \& Szalay, A. S. 2004, ApJ, 600, 681

\bibitem[Baldry et al. (2006)]{Baldry2006}Baldry, I. K., Balogh, M. L., Bower, R. G., Glazebrook, K., Nichol, R. C., Bamford, S. P., \& Budavari, T. 2006, MNRAS, 373, 469

\bibitem[Balogh et al. (1997)]{Balogh1997}Balogh, M. L., Morris, S. L., Yee, H. K. C., Carlberg, R. G., \& Ellingson, E. 1997, ApJ, 488, L75

\bibitem[Balogh \& Morris (2000)]{Balogh2000} Balogh, M. L., \& Morris, S. L. 2000, MNRAS, 318, 703

\bibitem[Barnes (1989)]{B89}Barnes, J. E. 1989, Nature, 338, 123

\bibitem[Barnes (1992)]{B92}Barnes, J. E. 1992, ApJ, 393, 484

\bibitem[Barton et al. (2000)]{Barton2000}Barton, E. J., Geller, M. J., \& Kenyon, S. J. 2000, ApJ, 530, 660

\bibitem[Bell et al (2004)]{B2004}Bell, E. F., et al. 2004, ApJ, 608, 752

\bibitem[Blaizot et al. (2005)]{B05} Blaizot J., Wadadekar Y., Guiderdoni B., Colombi S. T., Bertin E., Bouchet F. R., Devriendt J. E. G., Hatton S., 2005, MNRAS, 360, 159

\bibitem[Blanton et al. (2003)]{B2003} Blanton, M. R., et al. 2003, ApJ, 594, 186

\bibitem[Bower et al. (2006)]{B06} Bower, R. G., Benson, A. J., Malbon, R., Helly, J. C., Frenk, C. S., Baugh, C. M., Cole, S. \& Lacey, C. G. 2006, MNRAS, 370, 645

\bibitem[Bruzual \& Charlot (2003)]{BC2003} Bruzual G., \& Charlot S. 2003, MNRAS, 344, 1000

\bibitem[Butcher \& Oemler (1984)]{BO84} Butcher, H., \& Oemler, A., Jr. 1984, ApJ, 285, 426

\bibitem[Coziol \& Plauchu-Frayn (2007)]{CPF07} Coziol, R. \& Plauchu-Frayn, I. 2007, AJ, 133, 2630

\bibitem[Croton et al. (2006)]{C06} Croton, D. J., Springel, V.,
White, S. D. M., De Lucia, G., Frenk, C. S., Gao, L., Jenkins, A.,
Kauffmann, G., Navarro, J. F., \& Yoshida, N. 2006, MNRAS, 365, 11

\bibitem[Dekel \& Birnboim (2006)]{DB06} Dekel, A., \& Birnboim, Y. 2006, MNRAS, 368, 2

\bibitem[De Lucia \& Blaizot (2007)]{LB07} De Lucia, G. \& Blaizot, J. 2007, MNRAS, 375, 2

\bibitem[Deng et al. (2007)]{D07} Deng, X., He, J., Jiang, P., He, C., Luo, C.,  \& Wu, P. 2007. Astrophysics, 50, 18

\bibitem[Dressler (1980)]{D80} Dressler, A. 1980, ApJ, 236, 351

\bibitem[Dressler et al. (1997)]{D97} Dressler A. et al., 1997, ApJ, 490, 577

\bibitem[Ellison et al. (2008)]{Ellison2008}Ellison, S. L., Patton, D. R. Simard, L., \& McConnachie, A. W.  2008, AJ, 135, 1877

\bibitem[Faber et al. (2005)]{F05} Faber, S. M., et al. 2005, ApJ, submitted (astro-ph/0506044)

\bibitem[Farouki \& Shapiro (1982)]{FS82} Farouki, R. T., \& Shapiro, S. L. 1982, ApJ, 259, 103

\bibitem[Geller et al. (2006)]{Gell06} Geller, M. J., Kenyon, S. J., Barton, E. J., Jarrett, T. H., \& Kewley, L. J. 2006, AJ, 132, 2243

\bibitem[Gilbank \& Balogh (2008)]{GB2008} Gilbank D. G., Balogh M. L., 2008, MNRAS, 385, L116

\bibitem[Giovanelli \& Haynes (1983)]{GH83} Giovanelli R., \& Haynes M. P., 1983, AJ, 88, 881

\bibitem[Gisler (1980)]{G80}Gisler, G. 1980. A. J. 85, 623

\bibitem[Gladders \& Yee (2005)]{GY2005}Gladders, M. D., \& Yee, H. K. C. 2005, ApJS, 157, 1

\bibitem[G\'omez et al. (2003)]{Gomez2003} G\'omez, P. L., et al. 2003, ApJ, 584, 210


\bibitem[Gunn \& Gott (1972)]{GG72} Gunn, J. E., \& Gott, J. R. I. 1972, ApJ, 176, 1

\bibitem[Hashimoto et al. (1998)]{Hashimoto1998}Hashimoto, Y., Oemler, A., Jr., Lin, H., \& Tucker, D. L. 1998, ApJ, 499, 589

\bibitem[Haynes \& Giovanelli (1986)]{HG1986} Haynes, M.P., \& Giovanelli R. 1986, ApJ, 306, 466

\bibitem[Hernquist et al. (1995)]{Herquist95} Hernquist, L., Katz, N., \& Weinberg, D. H. 1995, ApJ, 442, 57

\bibitem[Hickson (1982), hereafter HCGs]{H82} Hickson, P. 1982, ApJ, 255, 382

\bibitem[Hickson et al. (1988)]{H88} Hickson, P., Kindl, E., \& Huchra, J. P. 1988, ApJ, 331, 64

\bibitem[Hickson (1990)]{H90} Hickson, P. 1990, in IAU Colloquium 124, Paired and Interacting Galaxies, edited by J. W. Sulentic, W.C. Keel
 \& C.M. Telesco, 77

\bibitem[Hickson et al. (1992)]{HMH92} Hickson, P., Mendes de Oliveira, C., Huchra, J. P., \& Palumbo, G. G. C. 1992, ApJ, 399, 353

\bibitem[Huchra \& Geller (1982)]{HG1982} Huchra J. P., \& Geller M. J., 1982, ApJ, 257, 423

\bibitem[Iglesias-P\'aramo \& V\'{i}chez (1999)]{IP1999} Iglesias-P\'aramo, J.,  \& V\'{i}chez, J. M.   1999, ApJ, 518, 94 

\bibitem[Kang \& van den Bosch (2008)]{Kang2008} Kang X., van den Bosch F. C., 2008, ApJ, 676, L101

\bibitem[Kauffmann et al. (1999)]{K99}Kauffmann G., Colberg J. M., Diaferio A., \& White S. D. M. 1999, MNRAS, 303, 188

\bibitem[Kennicutt (1983)]{K83}Kennicutt, R. C., Jr. 1983, AJ, 88, 483

\bibitem[Kennicutt  et al. (1987)]{K1987} Kennicutt, R. C., Keel, W. C., van der Hulst, J. M., Hummel, E., \& Roettiger, K. A. 1987, AJ, 93, 1011

\bibitem[Kitzbichler \& White (2008)]{M08} Kitzbichler, M.G.,\& White, S.D.M. 2008, MNRAS, submitted (arXiv:0804.1965)

\bibitem[Lambas et al. (2003)]{Lam03} Lambas, D. G., Tissera, P. B., Alonso, M. S., \& Coldwell, G. 2003, MNRAS, 346, 1189

\bibitem[Larson \& Tinsley (1978)]{LT1978}Larson, R. B., \& Tinsley, B. M. 1978, ApJ, 219, 46 


\bibitem[Larson, Tinsley \& Caldwell (1980)]{LTC80}Larson, R. B., Tinsley, B. M., \& Caldwell, C. N. 1980, ApJ, 237, 692

\bibitem[Lee et al. (2004)]{L04} Lee et al. 2004, ApJ, 127, 1811

\bibitem[Leon et al. (1998)]{L98} Leon, S., Combes, F., \& Menon, T. K. 1998, A\&A, 330, 37 
	
\bibitem[Lewis et al. (2002)]{Lewis2002} Lewis, I., et al. 2002, MNRAS, 334, 673

\bibitem[Li et al. (2007)]{Li2007} Li, C., Jing, Y. P., Kauffmann, G., B\"{o}rner, G., Kang, X., \& Wang, L. 2007, MNRAS, 376, 984

\bibitem[Maia et al. (1994)]{Maia1994} Maia, M. A. G., Pastoriza, M. G., Bica, E., \& Dottori, H. 1994, ApJS, 93, 425

\bibitem[McConnachie et al. (2008)]{M07} McConnachie, A., Ellison, S., \& Patton, D. 2008, MNRAS, in press (arXiv:0804.2928)

\bibitem[McCarthy et al. (2008)]{McC2008} McCarthy, I. G., et al. 2008, MNRAS, 383, 593

\bibitem[Mendes de Oliveira \& Hickson (1994)]{MH94} Mendes de Oliveira, C., \& Hickson, P. 1994, ApJ, 427, 684

\bibitem[Menon (1995)]{M95} Menon, T.K. 1995. MNRAS 274, 845

\bibitem[Nikolic et al. (2004)]{Nik04} Nikolic, B., Cullen, H., \& Alexander, P. 2004, MNRAS, 355, 874

\bibitem[Nilson (1973)]{N73} Nilson, P.N. 1973. Uppsala General Catalogue of Galaxies, Uppsala Obs. Ann. 6

\bibitem[Ostriker, Lubin, \& Hernquist (1995)]{OLH95} Ostriker, J. P., Lubin, L. M., \& Hernquist, L. 1995, ApJ, 444, L61

\bibitem[Palumbo et al. (1995)]{P95} Palumbo, G., Saracco, P., Hickson, P., \&
Mendes de Oliveira, C. 1995, ApJ, 109, 1476

\bibitem[Patiri et al. (1994)]{P94} Prandoni, I., Iovino, A., \& MacGillivray, H. T. 1994, AJ, 107, 1235

\bibitem[Patton et al. (2005)]{Pat05} Patton, D. R., Grant, J. K., Simard, L., Pritchet, C. J., Carlberg, R. G., \& Borne, K. D. 2005, AJ, 130, 2043

\bibitem[Prandoni et al. (2006)]{P06} Patiri, S., Prada F., Holtzman, J., Klypin, A., \& Betancort-Rijo,
J. 2006, MNRAS, 372, 1710

\bibitem[Poggianti et al. (1999)]{Poggianti1999}Poggianti, B. M., Smail, I., Dressler, A., Couch, W. J., Barger, A. J., Butcher, H., Ellis, R. S., \& Oemler, A. J. 1999, ApJ, 518, 576

\bibitem[Ponman et al. (1996)]{Ponman96} Ponman, T. J., Bourner, P. D. J., Ebeling, H., \& B\"{o}hringer, H. 1996, MNRAS, 283, 690

\bibitem[Postman et al. (2001)]{Postman2001} Postman, M., Lubin, L. M., \& Oke, J. B. 2001, AJ, 122, 1125

\bibitem[Rood \& Williams (1989)]{R89} Rood, H. J., \& Williams, B. A. 1989, ApJ, 339, 772

\bibitem[Roos \& Norman (1979)]{RN79} Roos, N., \& Norman, C.A., 1979, A\&A 76, 75

\bibitem[Schade et al. (1996)]{Sc1996}Schade, D., Carlberg, R. G., Yee, H. K. C., Lopez-Cruz, O., \& Ellingson, E. 1996, ApJ, 464, L63

\bibitem[Schwarzkopf \& Dettmar (2000)]{SD2000} Schwarzkopf, U., \& Dettmar, R. J. 2000, A\&AS, 144, 85 

\bibitem[Scodeggio \& Gavazzi (1993)]{SG93} Scodeggio, M., \& Gavazzi, G. 1993, ApJ, 409, 110

\bibitem[Stevens et al. (2004)]{S04} Stevens J. B., Webster R. L., Barnes D. G., Pisano D. J., \& Drinkwater M. J. 2004, P.A.S.A., 21, 318

\bibitem[Somerville, Primack, \& Faber (2001)]{SPF01} Somerville R. S., Primack J. R., \& Faber S. M. 2001, MNRAS, 320, 504

\bibitem[Springel et al. (2001)]{S2001} Springel V., White S. D. M., Tormen G., \& Kauffmann G. 2001, MNRAS, 328, 726

\bibitem[Springel et al. (2005)]{S2005} Springel et al. 2005, Nature, 435, 629.

\bibitem[Strateva et al. (2001)]{St2001} Strateva, I., et al. 2001. AJ, 122, 1861

\bibitem[Sulentic (1987)]{Sulentic1987}Sulentic, J. W. 1987, ApJ, 322, 605 

\bibitem[Sutherland \& Dopita (1993)]{SD93} Sutherland R. S. \& Dopita M. A., 1993, ApJS, 88, 253

\bibitem[Verdes-Montenegro et al. (2001)]{VM98}Verdes-Montenegro, L., Yun, M.S., Perea, J., del
  Olmo, A., \& Ho, P.T.P. 1998, ApJ, 497, 89

\bibitem[Verdes-Montenegro et al. (2001)]{VM01}Verdes-Montenegro, L., Yun, M. S., Williams, B. A., Huchtmeier, W. K., Del Olmo, A., \& Perea, J. 2001, A\&A, 377, 812

\bibitem[Weinmann et al. (2006)]{Wein2006} Weinmann, S. M., van den Bosch, F. C., Yang, X., Mo, H. J., Croton, D. J., \& Moore, B. 2006, MNRAS, 372, 1161

\bibitem[Whitmore, Gilmore, \& Jones (1993)]{W93} Whitmore, B. C., Gilmore, D. M., \& Jones, C. 1993, ApJ, 407, 489

\bibitem[Williams \& Rood (1987)]{WR87} Williams, B. A., \& Rood, H. J. 1987, ApJS, 63, 265 

\bibitem[Willmer et al. (2006)]{W2006}Willmer, C., et al. 2006, ApJ, 647, 853

\bibitem[Zabludoff et al. (1980)]{Z80} Zabludoff, A. I., Huchra, J. P., \&  Geller, M. J. 1990, ApJS, 74, 1

\bibitem[Zepf, Whitmore, \& Levison (1991)]{Z91} Zepf, S. E., Whitmore, B. C., \& Levison, H. F. 1991. ApJ, 383, 524 

\bibitem[Zheng et al. (1993)]{Z93} Zheng, J., Valtonen, M. J., \& Chernin, A. D. 1993, AJ, 105, 2047

\end{thebibliography}
\end{document}